\documentclass[twocol]{ametsoc}

\journal{jas}

\usepackage{ulem}
\usepackage{comment}
\usepackage{xcolor}
\bibpunct{(}{)}{;}{a}{}{,}

\ifdraft
\else
\usepackage{doi}
\usepackage{hyperref}
\hypersetup{
	breaklinks,
	colorlinks=true,
	linkcolor=blue,
	citecolor=purple,
	pdfstartview={FitH},
	pdfview={FitH 0},
	pdfauthor={N C Constantinou, B F Farrell and P J Ioannou},
	pdftitle={Statistical state dynamics of jet/wave coexistence in barotropic beta-plane turbulence},
 }
\fi

\def\Uv{\boldsymbol{U}}

\newcounter{saveeqn}%

\newcommand{\be}{\begin{equation}}
\newcommand{\ee}{\end{equation}}
\newcommand{\bdm}{\begin{equation*}}
\newcommand{\edm}{\end{equation*}}
\newcommand{\bea}{\begin{eqnarray}}
\newcommand{\eea}{\end{eqnarray}}

\newcommand{\partialf}[2]
{
 \ifthenelse{\equal{#1}{}}{\frac{\partial}{\partial #2}}{\frac{\partial #1}{\partial #2}}
}

\newcommand{\real}{\mathop{\mathrm{Re}}}
\newcommand{\imag}{\mathop{\mathrm{Im}}}

\renewcommand{\(}{\left(}
\renewcommand{\)}{\right)}
\renewcommand{\[}{\left[}
\renewcommand{\]}{\right]}
\newcommand{\<}{\left\langle}
\renewcommand{\>}{\right\rangle}

\newcommand{\Del}{\Delta}

\renewcommand{\d}{\delta}

\newcommand{\df}{\textrm{d}}

\newcommand{\s}{\sigma}
\renewcommand{\b}{\beta}
\renewcommand{\a}{\alpha}
\newcommand{\rM}{r_{\textrm{m}}}

\newcommand{\z}{\zeta}

\providecommand\bcdot{\boldsymbol{\cdot}}

\newsavebox{\astrutbox}
\sbox{\astrutbox}{\rule[-5pt]{0pt}{20pt}}

\newcommand{\e}{\varepsilon}

\def\bit{\vphantom{\dot{W}}}

\def\Rcal{\mathcal{R}}\def\Rcal{R}

\def\xv{\mathbf{x}}

\def\Uv{\mathbf{U}}
\def\uv{\mathbf{u}}

\def\kv{\mathbf{k}}

\def\dZ{\delta\tilde{Z}}
\def\dC{\delta\tilde{C}}

\def\ecz{\varepsilon_{c}}

\providecommand\bcdot{\boldsymbol{\cdot}}

\def\zhat{\hat{\mathbf{z}}}
\def\nablav{\bm\nabla}

\newcommand{\defn}{\ensuremath{\stackrel{\mathrm{def}}{=}}}
\renewcommand{\equiv}{\defn}

\usepackage{cancel}

\title{Statistical state dynamics of jet/wave coexistence in barotropic beta-plane turbulence}

\authors{Navid C. Constantinou\correspondingauthor{Navid Constantinou, University of Athens, Department of Physics, Section of Astrophysics, Astronomy and Mechanics, Building IV, Office 32, Panepistimiopolis, 15784 Zografos, Athens, Greece.}}
\affiliation{\small Cyprus Oceanography Center, University of Cyprus, Lefkosia, Cyprus}

\extraauthor{Brian F. Farrell}
\extraaffil{\small Department of Earth and Planetary Sciences, Harvard University, Cambridge, MA, USA}
 
\extraauthor{Petros J. Ioannou}
\extraaffil{\small Department of Physics, National and Kapodistrian University of Athens, Athens, Greece}

\email{navidcon@phys.uoa.gr}

\abstract{Jets coexist with planetary scale waves in the turbulence of planetary atmospheres. The coherent component of these structures arises from  cooperative interaction between the coherent structures and the incoherent small-scale turbulence in which they are embedded. It follows that  theoretical understanding of the dynamics of jets and planetary scale waves requires adopting the perspective of statistical state dynamics (SSD) which comprises the dynamics of the interaction between coherent and incoherent components in the turbulent state. In this work the S3T implementation of SSD for barotropic beta-plane turbulence is used to develop a theory for the jet/wave coexistence regime by separating the coherent motions consisting of the zonal jets together with a selection of large-scale waves from the smaller scale motions which  constitute the incoherent component. It is found that mean flow/turbulence interaction gives rise to jets that coexist with large-scale coherent waves in a synergistic manner. Large-scale waves that would exist only as damped modes in the laminar jet are found to be transformed into exponentially growing waves by interaction with the incoherent small scale turbulence which results in a change in the mode structure allowing the mode to tap the energy of the mean jet. This mechanism of destabilization differs fundamentally and serves to augment the more familiar S3T instabilities in which jets and waves arise from homogeneous turbulence with energy source exclusively from the incoherent eddy field and provides further insight into the cooperative dynamics of the jet/waves coexistence regime in planetary turbulence.
}

\begin{document}

\maketitle


\section{Introduction}

A regime in which jets, planetary scale waves and vortices coexist is
commonly observed in the turbulence of planetary atmospheres with the
banded winds and embedded vortices of Jupiter and the Saturn north polar
vortex constituting familiar
examples~\citep{Vasavada-and-Showman-05,Sanchez-Lavega-etal-2014}.
Planetary scale waves in the jet stream and vortices such as cutoff
lows are also commonly observed in the Earth's atmosphere. Conservation
of energy and enstrophy in undamped 2D turbulence implies continual transfer of
energy to the largest available spatial scales~\citep{Fjortoft-1953}.
This upscale transfer provides a conceptual basis for expecting the
largest scales to become increasingly dominant  as the energy
of turbulence forced at smaller scale is continually transferred to the
larger scales. However, the observed large-scale structure in planetary
atmospheres is dominated not by incoherent large-scale turbulent motion
as would be expected to result from the incoherent phase relation of
Fourier modes in a turbulent cascade, but rather by coherent zonal jets,
vortices and waves of highly specific form. Moreover, the scale of these
coherent structures is distinct from the largest scale permitted in the
flow. An early attempt to understand the formation of jets in planetary
turbulence did not address the structure of the jet beyond attributing
the jet scale to arrest of the incoherent upscale energy cascade at the
length scale set by the value of the planetary vorticity gradient and a
characteristic flow velocity~\citep{Rhines-1975}. In Rhines's
interpretation this is the scale at which the turbulent energy cascade
is intercepted by the formation of propagating Rossby waves. 
While this result provides a
conceptual basis for expecting zonal structures with spatial scale
limited by the planetary vorticity gradient to form in beta-plane
turbulence, the physical mechanism of formation, the precise morphology
of the coherent structures and their stability are not addressed by
these general considerations.

Our goal in this work is to continue development of a general theory 
for the formation of finite amplitude structures
in planetary turbulence, specifically addressing the regime in which jets and planetary waves coexist. 
This theory identifies specific mechanisms responsible for formation and equilibration of coherent structures in planetary turbulence.
A number of mechanisms have been previously advanced to account for jet, wave and vortex formation. One
such mechanism which addresses exclusively jet formation is vorticity mixing 
by breaking Rossby waves  leading to homogenization of potential vorticity (PV) in
localized regions \citep{Baldwin-etal-2007,Dritschel-McIntyre-2008}
resulting in the case of barotropic beta-plane turbulence in broad
retrograde parabolic jets and relatively narrow prograde jets with
associated staircase structure in the absolute vorticity. While PV
staircases have been obtained in some numerical simulations of strong
jets~\citep{Scott-Dritschel-2012}, vorticity mixing in the case of weak
to moderately strong jets is insufficient to produce a prominent
staircase structure. Moreover, jets have been shown to form as a
bifurcation from homogeneous turbulence in which case the jet is
perturbative in amplitude and wave breaking is not
involved~\citep{Farrell-Ioannou-2003-structural,Farrell-Ioannou-2007-structure}.

Equilibrium statistical mechanics has also been advanced to explain
formation of coherent structures e.g.~by~\cite{Miller-1990}
and~\cite{Robert-Sommeria-1991}. The principle is that dissipationless
turbulence tends to produce configurations that maximize entropy while
conserving both energy and enstrophy. These maximum entropy
configurations in beta-plane turbulence assume the coarse grained
structure of zonal jets  (cf.~\cite{Bouchet-Venaile-2012}).
However, the relevance of these results to the formation, equilibration
and maintenance of jets in strongly forced and dissipated
planetary flows remains to be established.

Zonal jets and waves can also arise from modulational
instability~\citep{Lorenz-1972, Gill-1974, Manfroi-Young-99,
Berloff-etal-2009a, Connaughton-etal-2010}. This instability produces
spectrally non-local transfer to the unstable structure from forced waves and
therefore presumes a continual source of waves with the required form.
In baroclinic flows, baroclinic instability has been advanced as the
source of these waves~\citep{Berloff-etal-2009a}. From the broader
perspective of the statistical state dynamics theory used in this work, 
modulational instability is a special case of an S3T structural 
instability~\citep{Parker-Krommes-2014-generation,Parker-2014,Bakas-etal-2015}. 
However, modulational instability does not include the mechanisms for realistic equilibration of the instabilities at finite amplitude although a Landau-type term has been used to produce equilibration of the modulational instability (cf.~\cite{Manfroi-Young-99}).

Another approach to understanding the jet/wave coexistence regime is based on the idea that jets and waves interact in a cooperative manner. Such a dynamic is suggested for example by observations of a prominent \mbox{wavenumber-5} disturbance in the Southern Hemisphere~\citep{Salby-1982}. Using a zonally symmetric two-layer baroclinic model \citet{Cai-Mak-1990} demonstrated that storm track organization by a propagating planetary scale wave resulted in modulation in the distribution of synoptic scale transients configured on average  to maintain the organizing planetary scale wave. 
The symbiotic forcing by synoptic scale transients which on 
average maintains planetary scale waves was traced to barotropic interactions in the studies of \citet{Robinson-1991b} and \citet{Qin-Robinson-1992}. While diagnostics of simulations such as these  are suggestive,
comprehensive analysis of the essentially statistical mechanism of the symbiotic regime requires obtaining solutions 
of the statistical state dynamics underlying  it  and indeed  the present
work identifies an underlying statistical mechanism
by which transients are systematically
organized by a planetary scale wave so as
to on average support that planetary scale wave in a spectrally
non-local manner.

Stochastic structural stability theory (S3T) provides a statistical
state dynamics (SSD) based theory accounting for the formation, equilibration and
stability of coherent structures in turbulent flows. The underlying
mechanism of jet and wave formation revealed by S3T is the 
spectrally non-local interaction between the large-scale structure and
the small-scale turbulence~\citep{Farrell-Ioannou-2003-structural}. S3T
is a non-equilibrium statistical theory based on a closure comprising
the nonlinear dynamics of the coherent large-scale structure together with the
consistent second-order fluxes arising from the  incoherent eddies. The S3T system is a
cumulant expansion of the turbulent dynamics closed at second 
order~(cf.~\cite{Marston-etal-2008}), which has been shown
to become asymptotically exact for large-scale jet dynamics in turbulent flows in the limit of zero forcing and dissipation and  infinite 
separation between the time scales
of evolution of the large-scale jets and the eddies \citep{Bouchet-etal-2013,Tangarife-2015-phd}.
S3T has been employed to understand the 
emergence and equilibration of zonal jets in planetary turbulence
in barotropic flows on a beta-plane and on the sphere \citep{Farrell-Ioannou-2003-structural,
Farrell-Ioannou-2007-structure,Farrell-Ioannou-2009-equatorial,Marston-etal-2008,Srinivasan-Young-2012, Marston-2012,Constantinou-etal-2014,Bakas-Ioannou-2013-jas,Tobias-Marston-2013,
 Parker-Krommes-2014-generation, Ait-Chaalal-etal-2015}, in baroclinic two layer turbulence \citep{Farrell-Ioannou-2008-baroclinic,Farrell-Ioannou-2009-closure} and
 in drift-wave turbulence in plasmas \citep{Farrell-Ioannou-2009-plasmas,Parker-Krommes-2013}.
It has been used in order to study the emergence and equilibration of finite amplitude propagating
non-zonal structures in barotropic flows \citep{Bakas-Ioannou-2013-prl,
Bakas-Ioannou-2014-jfm,Bakas-etal-2015} and the dynamics of blocking in two-layer baroclinic atmospheres \citep{Bernstein-Farrell-2010}. It has also been used to study the role of coherent structures in the dynamics of the 3D turbulence of 
wall-bounded shear flows \citep{Farrell-Ioannou-2012,Thomas-etal-2014-sustain,Thomas-etal-2015-minimal,Farrell-etal-2015-vlsm}.

In certain cases a barotropic S3T homogeneous turbulent equilibrium 
undergoes a bifurcation in which non-zonal coherent structures emerge as a function of turbulence 
intensity prior to the emergence of zonal jets and  when zonal jets emerge a new type of 
jet/wave equilibrium forms \citep{Bakas-Ioannou-2014-jfm}. 
In this paper we use S3T to further examine the dynamics of the jet/wave
coexistence regime in barotropic beta-plane turbulence. In order to probe the jet/wave/turbulence dynamics in 
more depth a separation is made between the coherent jets and 
large-scale waves and the smaller scale motions which are considered to 
constitute the incoherent turbulent component of the flow. 
This separation is accomplished using a dynamically consistent projection in 
Fourier space.  By this means we show
 that jet states maintained by turbulence 
 may be unstable to emergence of non-zonal traveling waves and trace
 these unstable eigenmodes to what would, in the absence of turbulent fluxes,
 have been damped wave modes of the mean
jet. Thus we show that the cooperative dynamics between large-scale
coherent and small-scale incoherent motion
is able to transform damped modes into unstable modes by
altering the mode structure
allowing it to tap the energy of the mean jet. 

 In this work we also extend the S3T stability analysis of homogeneous equilibria \citep{Farrell-Ioannou-2003-structural,Farrell-Ioannou-2007-structure,Srinivasan-Young-2012,Bakas-Ioannou-2013-prl,Bakas-Ioannou-2013-jas,Bakas-Ioannou-2014-jfm,Bakas-etal-2015} to the S3T stability of jet equilibria. We present new methods for the calculation of the S3T stability of jet equilibria, extending the work of \cite{Farrell-Ioannou-2003-structural, Parker-Krommes-2014-generation}, which was limited to the study of the S3T stability of jets only with respect to zonal perturbations, to the S3T stability of jets to non-zonal perturbations.

\section{Formulation of  S3T dynamics for barotropic \mbox{$\beta$-plane} turbulence}

Consider a non-divergent flow $\uv=(u,v)$ on a $\beta$-plane with
coordinates $\xv=(x,y)$; in which $x$ is the zonal direction and  $y$ the
meridional direction, and with the flow confined to a periodic
channel of size $2 \pi L \times 2\pi L$. The velocity field  can be obtained from
a streamfunction $\psi$ as 
$\uv=\zhat\times\nablav\psi$, with $\zhat$ the unit vector normal to the $(x,y)$ plane.
The component of vorticity normal to the plane of
motion is $\z\equiv \partial_x v - \partial_y u$ is given as $\z=\Del\psi$ with $\Delta \equiv\partial^2_{x}+\partial^2_{y}$ the Laplacian 
operator. In the presence of dissipation and stochastic excitation the vorticity evolves according to:
\be
\partial_t\z = -\uv\bcdot\nablav \z - \b v - r \z + \nu \Del \z + \sqrt{\e}\xi\ ,\label{eq:NLbarotropic}
\ee
in which the flow is damped by Rayleigh dissipation with coefficient $r$ and viscous dissipation with
coefficient $\nu$. 
The stochastic excitation maintaining the turbulence, $\xi(\xv,t)$, is a  Gaussian random process that 
is temporally delta-correlated with zero mean.

Equation~\eqref{eq:NLbarotropic} is non-dimensionalized  using length scale $L$ and  time scale $T$. The double periodic domain
becomes $2\pi\times2\pi$ and the non-dimensional variables in~\eqref{eq:NLbarotropic} are: 
$\z^*=\z/T^{-1}$, $\uv^*=\uv/(LT^{-1})$, $\xi^*=\xi/(L^{-1}T^{-1/2})$, $\e^*=\e/(L^2T^{-3})$, $\b^*=\b/(LT)^{-1}$, $r^*=r/T^{-1}$ and $\nu^*=\nu/(L^2T^{-1})$, where asterisks denote non-dimensional units. Hereafter all variables are assumed non-dimensional and the asterisk is omitted.

 We  review now the formulation of the S3T approximation to the statistical state dynamics (SSD) of~\eqref{eq:NLbarotropic}.
The S3T dynamics was introduced  in the matrix formulation by \cite{Farrell-Ioannou-2003-structural}.
 \cite{Marston-etal-2008} showed that S3T comprises a canonical  second-order closure of the exact statistical state dynamics and derived it  alternatively
 using  the Hopf formulation. \cite{Srinivasan-Young-2012} obtained a continuous formulation which 
 facilitates analytical explorations of S3T stability of turbulent statistical equilibria.
%
 

An averaging operator by which mean quantities are obtained, denoted by angle brackets,
$\langle \;\bcdot\;\rangle$, is required in order to form the S3T equations. Using this averaging operator
 the flow streamfunction is decomposed as
\be
\z = Z +\z'\ ,
\ee
where  
\be
Z(\xv,t)\equiv\<\z(\xv,t)\>~,
\ee
is the mean field or the first cumulant of the vorticity and, similarly, for the derived flow fields, i.e. $\Uv$, $\psi$. The  eddies, $\z'$,  satisfy  the important property that 
\be
\< \z' \>=0\ ,\label{eq:fl0}
\ee
which relies on the averaging operation satisfying the Reynolds condition
(cf. \cite{Ait-Chaalal-etal-2015}) that for any two fields $f$ and $g$,
\be
\< \bit \< f \> g \> = \< f \> \< g \>~.\label{eq:Rp1}
\ee

The equation for the first cumulant  
is obtained by averaging equation~\eqref{eq:NLbarotropic}, which after repeated use of~\eqref{eq:Rp1} becomes:
\ifdraft
\begin{align}
\partial_t Z + \Uv \bcdot \nablav  Z  +\b V + r Z - \nu \Del Z &= -  \nablav \bcdot \left < \uv' \zeta' \right  >  \ , \label{eq:MNL} 
\end{align}
\else
\begin{align}
\partial_t Z + \Uv \bcdot \nablav  Z  +\b V + r Z - \nu \Del Z &= -  \nablav \bcdot \left < \uv' \zeta' \right  >   \ , \label{eq:MNL} 
\end{align}\fi
in which we have we assumed $\left < \, \xi \,\right > = 0$.  The term \mbox{$-\nablav \bcdot \< \uv' \z' \>$}  represents  the source of
 mean vorticity arising from the perturbation vorticity flux divergence.

The second cumulant of the  vorticity fluctuations is the covariance
\be
C(\xv_a,\xv_b, t)\equiv \left<\zeta'(\xv_a,t)  \zeta'(\xv_b,t) \right>\ ,\label{eq:C}
\ee
which  is a function of five variables: time, $t$, and the coordinates of  the two points $\xv_a$ and $\xv_b$. 
We write~\eqref{eq:C} concisely as $C_{ab}= \left<\zeta_a'  \zeta_b' \right>$.

All second  moments of the velocities  can be expressed as  linear functions of $C$. 
For example the perturbation vorticity flux divergence source term, 
$\nablav\bcdot  \left < \uv'(\xv,t) \z'(\xv,t) \right >$,
in the mean vorticity equation~\eqref{eq:MNL}
can be written  as a function of $C$ as follows: 
%
\begin{align}
 \nablav \bcdot \left < \uv' \zeta' \right > & = \frac1{2} \nablav\bcdot\< \uv_a' \z_b'  + \uv_b'  \z_a' \>_{a=b}\nonumber\\
  &= \frac{1}{2}\nablav\bcdot\[\zhat\times\(\nablav_a\Del_a^{-1}+\nablav_b\Del_b^{-1}\) \left < \z_a' \z_b' \right >  \]_{a=b} \nonumber\\
 &= \frac{1}{2}\nablav\bcdot\[\zhat \times\(\nablav_a\Del_a^{-1}+\nablav_b\Del_b^{-1}\)C_{ab}\]_{a=b}\nonumber \\
 &\equiv R(C) \, \label{eq:uv}
\end{align}   
in which $\uv'_j\equiv\uv'(\xv_j,t)$ and subscripts in the differential operators indicate 
the specific independent spatial variable  the operator is defined on. To derive~\eqref{eq:uv} we used that  $\uv_j'=\zhat\times\nablav_j \Del_j^{-1}\z_j'$, with $\Del^{-1}$  the inverse Laplacian.
The notation $a=b$ indicates that  the function  of the five independent variables, $\xv_a$,
$\xv_b$  and $t$, in~\eqref{eq:uv} is to be considered a function of  two independent spatial variables and $t$
 by setting
$\xv_a=\xv_b=\xv$. By denoting 
the divergence of the mean of the perturbation vorticity flux
$\nablav \bcdot \left < \uv' \zeta' \right >$ in~\eqref{eq:uv} as 
$R(C)$  we underline that the  forcing of the mean vorticity equation~\eqref{eq:MNL} by the perturbations depends on the second cumulant (the covariance of
the vorticity field).  Adopting this notation for the divergence of the mean of the perturbation vorticity flux, the equation for the mean vorticity (the first cumulant)~\eqref{eq:MNL} takes the form:
\begin{subequations}
\ifdraft
\begin{align}
\partial_t Z   + \Uv \bcdot\nablav Z    +\b V +r Z - \nu \Del Z &= - R(C)\ , \label{eq:MNLC} 
\end{align}
\else
\begin{align}
\partial_t Z   + \Uv \bcdot\nablav Z 
+\b V +r Z - \nu \Del Z &= -R(C)\ . \label{eq:MNLC} 
\end{align}\fi
The equation for the perturbation vorticity is obtained by 
subtracting~\eqref{eq:MNL} from~\eqref{eq:NLbarotropic}:
\ifdraft
\begin{align}
\partial_t \zeta' &= -  \( \Uv  \bcdot \nablav \z'  + \uv' \bcdot \nablav  Z \) -\left ( \uv' \bcdot \nablav \zeta' -  \left < \uv' \bcdot \nablav \zeta'   \right > \right )-\b v'- r \z' +\nu \Del \z' + \sqrt{\e}\xi  \ ,
\label{eq:FNL}
\end{align}\else
\begin{align}
\partial_t \zeta' &= -  \( \Uv  \bcdot \nablav \z'  + \uv' \bcdot \nablav  Z \) - \nablav\bcdot \left ( \uv' \zeta' -  \left < \uv'  \zeta'   \right > \right ) \nonumber  \\
& \hspace{3em} -\b v'- r \z' +\nu \Del \z' + \sqrt{\e}\xi  \nonumber \\
&= A \zeta' - \nablav\bcdot \left ( \uv' \zeta' -  \left < \uv'  \zeta'   \right > \right ) + \sqrt{\e}\xi  \ ,
\label{eq:FNL}
\end{align}
\fi\end{subequations}
where 
\be
A\equiv-\Uv\bcdot\nablav - \[\bit\beta\partial_{x} - (\Delta\Uv)\bcdot\nablav \]\Del^{-1} -   r  + \nu \Del \ .\ \label{eq:op_A}
\ee
Using~\eqref{eq:FNL}, definition~\eqref{eq:C} and noting that $\<\z'\>=0$ we obtain the evolution equation 
for $C$:
%
\ifdraft
\begin{align}
\partial_t C &= \left < \z'_a \partial_t \z_b' + \z'_b \partial_t \z_a' \right > \nonumber \\
&= (A_a + A_b) C + \sqrt{\e}\<\z'_a\xi_b + \z'_b\xi_a\>  +  \<\bit  \[ \nablav_a \bcdot \( \uv'_a   \z'_a \)\bit\]\z'_b +  \[ \nablav_b \bcdot \( \uv'_b   \z'_b \)\bit\] \z'_a\>\ .\label{eq:C_v1}
\end{align}
\else
\begin{align}
\partial_t C &= \left < \z'_a \partial_t \z_b' + \z'_b \partial_t \z_a' \right > \nonumber \\
&=
(A_a + A_b) C + \sqrt{\e}\<\z'_a\xi_b + \z'_b\xi_a\>   \nonumber\\
&\qquad +  \<\bit  \[ \nablav_a \bcdot \( \uv'_a   \z'_a \)\bit\]\z'_b +  \[ \nablav_b \bcdot \( \uv'_b   \z'_b \)\bit\] \z'_a\>\ .\label{eq:C_v1}
\end{align}\fi
Both terms $\langle  \[ \nablav_a \bcdot \( \uv'_a   \z'_a \)\bit\]\z'_b \rangle$ and $\langle  \[ \nablav_b \bcdot \( \uv'_b   \z'_b \)\bit\] \z'_a \rangle$ in~\eqref{eq:C_v1} can be expressed as  linear functions of the third cumulant of the vorticity fluctuations, $\Gamma_{abc}\equiv\<\z'_a\z'_b\z'_c\>$, e.g.\ifdraft
\begin{align}
\left\langle  \[ \nablav_a \bcdot \( \uv'_a   \z'_a \)\bit\]\z'_b \right\rangle &=\frac1{2}\left\langle   \nablav_a \bcdot \[ \uv'_a  \z'_c+\uv'_c  \z'_a \]_{c\to a}\,\z'_b \right\rangle \nonumber\\
&= \frac1{2}  \nablav_a \bcdot\[ \zhat\times \( \nablav_a\Del^{-1}_a+\nablav_c\Del^{-1}_c\) \left <  \z'_a \z'_b\z'_c \right >  \bit\]_{c\to a}\nonumber\\
&=   \frac1{2} \nablav_a \bcdot\[\zhat\times\( \nablav_a\Del^{-1}_a+\nablav_c\Del^{-1}_c\) \Gamma_{abc}  \bit\]_{c\to a}\ ,
\label{eq:11}
\end{align}\else
\begin{align}
&\left\langle  \[ \nablav_a \bcdot \( \uv'_a   \z'_a \)\bit\]\z'_b \right\rangle =\nonumber\\
&\hspace{1em}=\frac1{2}\left\langle   \nablav_a \bcdot \[ \uv'_a  \z'_c+\uv'_c  \z'_a \]_{c\to a}\,\z'_b \right\rangle \nonumber\\
&\hspace{1em}= \frac1{2}  \nablav_a \bcdot\[ \zhat\times \( \nablav_a\Del^{-1}_a+\nablav_c\Del^{-1}_c\) \left <  \z'_a \z'_b\z'_c \right >  \bit\]_{c\to a}\nonumber\\
&\hspace{1em}=   \frac1{2} \nablav_a \bcdot\[\zhat\times\( \nablav_a\Del^{-1}_a+\nablav_c\Del^{-1}_c\) \Gamma_{abc}  \bit\]_{c\to a}\ ,
\label{eq:11}
\end{align}\fi
explicitly revealing that  the dynamics of the second cumulant of the vorticity fluctuations 
is not closed. Notation $c\to a$ indicates that the function of independent
spatial variables $\xv_a$, $\xv_b$ and $\xv_c$ should be considered a 
function of only $\xv_a$ and $\xv_b$ after setting 
$\xv_c\to\xv_a$.  The S3T system is obtained by truncating the cumulant 
expansion at second order either by setting the third cumulant term in~\eqref{eq:C_v1} equal to zero or by 
assuming that the third cumulant term  is proportional  to  a state independent  
covariance $Q(\xv_a,\xv_b)$.  The latter is equivalent to representing both  
the nonlinearity, $\nablav\bcdot \( \uv' \z' - \< \uv' \z' \> \)$,   
and the externally imposed stochastic excitation together as a single stochastic excitation  $\sqrt{\e}\xi(\xv,t)$  with zero mean 
and  two point and two time correlation function:
 \be
 \left < \xi(\xv_a,t_1) \xi(\xv_b,t_2) \right > = \delta(t_1-t_2)\, Q(\xv_a,\xv_b)\ ,\label{eq:xi_cor}
 \ee
from which  it can be shown that\footnote{Assumption~\eqref{eq:xi_cor} implies identity~\eqref{eq:xi_zeta}  even when $\z'$ obeys the nonlinear~\eqref{eq:FNL} (cf.~\cite{Farrell-Ioannou-2015-book,Constantinou-2015-phd}.}:
\be
\<\z'_a\xi_b+ \z'_b \xi_a\> = \sqrt{\e} Q(\xv_a,\xv_b)\ ,\label{eq:xi_zeta}
\ee
and consequently~\eqref{eq:C_v1} simplifies to the time dependent Lyapunov equation:
\begin{equation}
\partial_t C = (A_a + A_b) C + \e Q\ ,
\end{equation}
in which  the subscripts $a$, $b$ on $C$ and $Q$ are implied.

Using parametrization~\eqref{eq:xi_cor} to account for both the perturbation nonlinearity, 
$\nablav\bcdot \left ( \uv' \zeta' -  \left < \uv'  \zeta'   \right > \right )$, and the external 
stochastic excitation, $\sqrt{\e} \xi$
implies that full correspondence  between the  mean equation~\eqref{eq:MNLC} coupled with 
the parameterized perturbation equation~\eqref{eq:FNL} and  the nonlinear dynamics~\eqref{eq:NLbarotropic}  
requires that the stochastic term  accounts fully for modification of the 
 perturbation spectrum  
 by the perturbation 
 nonlinearity in addition to the explicit externally imposed stochastic excitation. 
 It follows that  the stochastic parameterization   required to obtain
 agreement between the approximate statistical state dynamics  and the nonlinear simulations 
 differs from the explicit external forcing alone unless 
  the eddy--eddy interactions are negligible.

The resulting S3T system is an autonomous dynamical system involving only the first two cumulants 
that determines  their consistent evolution. The S3T system for a 
chosen averaging operator is:\begin{subequations}\label{eq:s3ts}\begin{align}
&\partial_t Z=-\mathbf{U}\bcdot \nablav Z - \b V-  r Z + \nu \Del Z-R(C)\ \label{eq:s3tm} ,\\
&\partial_t C=(A_a+A_b)C+\e\,Q\ .\label{eq:cov_evo}
\end{align}\label{eq:s3t}\end{subequations}

%


For the purpose of  studying  turbulence dynamics  it is appropriate to choose an averaging operator that isolates the physical mechanism of interest.  
Typically the averaging operator is chosen to separate the coherent structures  from the incoherent turbulent 
motions.  Coherent structures are critical components of turbulence in shear flow both in the 
energetics of interaction between the large and small scales and in the mechanism by which the 
statistical steady state is determined.  
Retaining  the nonlinearity and structure of these flow components is crucial to constructing a theory of 
shear flow turbulence that properly  accounts for the role of the coherent structures.  In contrast, 
nonlinearity and detailed structure  information is  not required to account for the role of the 
incoherent motions and the statistical information contained in the second cumulant suffices to 
include the influence of these on the turbulence  dynamics. This results in a great practical as 
well as conceptual simplification that allows a theory of turbulence to be constructed. 
In the case of  beta-plane turbulence a phenomenon of interest is the formation of 
coherent zonal jets from the background of incoherent turbulence.  To isolate the dynamics of  jet formation  
zonal averaging is appropriate. 
Alternatively,  if the focus of study is  the emergence of large planetary scale  waves
the averaging operation  would be an appropriate extension of the 
Reynolds average over an intermediate spatial scale to produce a spatially   coarse grained /fine grained flow separation.
An averaging operation of this form was used  by \citep{Bernstein-Farrell-2010}
in   their S3T study  of blocking  in a two-layer baroclinic atmosphere and by
\cite{Bakas-Ioannou-2013-prl,Bakas-Ioannou-2014-jfm} to provide an explanation for the emergence of 
travelling wave structures (``zonons") in barotropic turbulence. However, the Reynolds average
defined over an intermediate time or spatial scale:\begin{subequations}
\begin{align}
\left < f (\xv,t) \right > &\equiv \frac{1}{2T} \int_{t-T}^{t+T}\df \tau\; f(\xv, \tau) \ ,\\
&{\rm or}~~~\nonumber \\ 
\left < f (\xv,t) \right > &\equiv \frac{1}{4 X Y} \int_{x-X}^{x+X} \df x' \int_{y-Y}^{y+Y} \df y'\;f(\xv', t)\ ,
\end{align}\end{subequations}
satisfies the Reynolds condition~\eqref{eq:Rp1} only  approximately and  to the extent that  there is adequate scale separation.
The S3T system  that was derived in~\eqref{eq:s3ts} is exact if the averaging operation is the zonal average and an adequate 
approximation for jets and a 
selection of large-scale waves if there is sufficient scale separation to satisfy the Reynolds condition~\eqref{eq:Rp1}.

Because the scale separation assumed in~\eqref{eq:s3ts}
is only approximately satisfied in
many cases of interest,  an alternative formulation of S3T  will now be obtained  in 
which separation  into two independent
interacting components of different scales is implemented   (a similar formulation is used by \cite{Marston-etal-2016}).
This formulation makes more precise
the dynamics of the coherent jet and wave interacting with incoherent turbulence regime in S3T.
 
The required separation is obtained by projecting the dynamics~\eqref{eq:NLbarotropic} on two  
distinct sets of Fourier harmonics. Consider the Fourier expansion of
the streamfunction,
\be
\psi= \sum_{k_x} \sum_{k_y} \hat{\psi}_{\kv} e^{i \kv \bcdot \xv}~,
\ee
with  $\kv=(k_x,k_y)$ and the projection operator  $P_K$ defined as (cf.~\cite{Frisch-1995}):
\be
P_K \psi \equiv \sum_{|k_x| \le K} \sum_{k_y}  \hat{\psi}_{\kv} e^{i \kv \bcdot \xv}\ ,
\ee
so that the large-scale flow is identified through streamfunction $\Psi=P_K \psi$ and the small-scale flow through $\psi' = (I-P_K) \psi$ where:
\be
(I-P_K) \psi \equiv \sum_{|k_x| > K} \sum_{k_y} \hat{\psi}_{\kv} e^{i \kv \bcdot \xv }\ ,
\ee
with $I$ the identity. 
Similarly,  vorticity  and velocity fields are decomposed into $\zeta = Z + \zeta'$ and $\uv = \Uv + \uv'$.

From~\eqref{eq:NLbarotropic} and under the assumption that
the stochastic excitation projects only on the small scales,  the large scales evolve according to:\begin{subequations}
\ifdraft
\begin{align}
\partial_t Z  =&   -P_K \[ \Uv \bcdot\nablav Z+ \nablav \bcdot \(\uv'  \zeta' \)\] -P_K\( \Uv \bcdot \nablav \zeta'  + \uv' \bcdot \nablav Z \) -\b V- r Z + \nu \Del Z\ , \label{eq:PKMNL} 
\end{align}
\else
\begin{align}
\partial_t Z  =&   -P_K \[ \Uv \bcdot\nablav Z+ \nablav \bcdot \(\uv'  \zeta' \)\]  \nonumber\\
&  -P_K\( \Uv \bcdot \nablav \zeta'  + \uv' \bcdot \nablav Z \)\nonumber\\
& -\b V- r Z + \nu \Del Z \ , \label{eq:PKMNL} 
\end{align}
\fi
while  the small scales evolve  according to:
\ifdraft
\begin{align}
\partial_t \zeta'  = &-\left (I -P_K \right )  \left ( \Uv  \bcdot \nablav \zeta'  +
\uv' \bcdot \nablav Z  \right ) - \left ( I  - P_K \right ) \[ \Uv \bcdot \nablav Z   +
  \nablav \bcdot\(\uv'\zeta' \)\] -\b v'- r \zeta' +\nu \Del \zeta' + \sqrt{\e}\xi \ .\label{eq:PKPNL}
\end{align}
\else
\begin{align}
\partial_t \zeta'  = &-\left (I -P_K \right )  \left ( \Uv  \bcdot \nablav \zeta'  +
\uv' \bcdot \nablav Z  \right ) \nonumber \\ 
&- \left ( I  - P_K \right ) \[ \Uv \bcdot \nablav Z   +
  \nablav \bcdot\(\uv'\zeta' \)\] \nonumber \\
&-\b v'- r \zeta' +\nu \Del \zeta' + \sqrt{\e}\xi \ .\label{eq:PKPNL}
\end{align}\fi\label{eq:PKNL}\end{subequations}
If $P_K$ were an averaging operator that satisfied  the Reynolds condition~\eqref{eq:Rp1}
term $P_K \( \Uv \bcdot \nablav\z'  + \uv' \bcdot \nablav Z\)$ in~\eqref{eq:PKMNL} would vanish. 
Here it  does not, as both of these terms scatter energy to the large scales. However, an energetically closed
S3T system for the first two cumulants can be derived
by making the quasi-linear approximation
(QL) in~\eqref{eq:PKPNL}, i.e.~neglect the terms $\left ( I  - P_K \right ) \[ \Uv \bcdot \nablav Z   +
  \nablav \bcdot\(\uv'\zeta' \)\] $ that represent
 projection of the eddy--eddy and large-scale--large-scale
interactions on the eddy flow components, and additionally neglect the
terms $P_K( \Uv \bcdot \nablav \zeta'  + \uv' \bcdot \nablav Z ) $ in the large-scale equation~\eqref{eq:PKMNL}.
These later terms  as well as $\(I - P_K \)\(\Uv\bcdot\nablav Z\)$ are not of primary importance to the dynamics and in any case vanish with sufficient
scale separation.
With these terms neglected we obtain the  projected QL system:\ifdraft\begin{subequations}
\begin{align}
\partial_t Z  =&  -\b V- r Z + \nu \Del Z -P_K \[ \Uv \bcdot\nablav Z+ \nablav \bcdot \(\uv'  \zeta' \)\]  \ ,\label{eq:MQL} \\
\partial_t \zeta' = &-\(I -P_K \right )   \( \Uv  \bcdot \nablav \z'  + \uv' \bcdot \nablav  Z \)-\b v'- r \z' +\nu \Del \z' + \sqrt{\e}\xi \ ,\label{eq:PQL}
\end{align}\label{eq:PKQL}\end{subequations}
\else
\begin{subequations}
\begin{align}
\partial_t Z  =&  -\b V- r Z + \nu \Del Z -P_K \[ \Uv \bcdot\nablav Z+ \nablav \bcdot \(\uv'  \zeta' \)\]  \ ,\label{eq:MQL} \\
\partial_t \zeta' = &-\(I -P_K \right )   \( \Uv  \bcdot \nablav \z'  + \uv' \bcdot \nablav  Z \) \nonumber \\
&\qquad\qquad-\b v'- r \z' +\nu \Del \z' + \sqrt{\e}\xi \ ,\label{eq:PQL}
\end{align}\label{eq:PKQL}\end{subequations}
\fi
which  conserves   total energy and enstrophy in the absence of forcing and dissipation. The conservation properties of the full 
barotropic equations are
retained because the typically small terms that have been discarded scatter energy and enstrophy between~\eqref{eq:PKNL}.

Assuming $Z=P_K(\zeta)$ is the coherent flow and $C= \left < \zeta' (\xv_a,t) \zeta'(\xv_b,t) \right >$  the covariance  of 
the incoherent eddies, with $\langle \;\bcdot\;\rangle$ being an average over forcing realizations, we
obtain the  corresponding S3T system for the first two cumulants:
\begin{subequations}\label{eq:s3tf}\begin{align}
& \partial_t Z=- \b V -r Z + \nu \Del Z-P_K \[\bit \mathbf{U}\bcdot \nablav Z +R(C) \] \ ,\label{eq:s3tmf}\\
&\partial_t C= (I-P_{K a})  A_a\, C +  (I-P_{K b}) A_b \,C  +\e\,Q\ .\label{eq:cov_evof}
\end{align}\label{eq:s3t_Pk}\end{subequations}
It can be shown that these equations have the same quadratic conservation properties as the S3T equations~\eqref{eq:s3ts} and the full nonlinear equations~\eqref{eq:NLbarotropic}. Note that for $K=0$ this projection formulation reduces to the zonal mean/perturbation formulation employed previously to study zonal jet formation~\citep{Farrell-Ioannou-2003-structural, Farrell-Ioannou-2007-structure,Srinivasan-Young-2012}.

\section{Specification of the parameters used in this work}

Assume that the large-scale phase coherent motions occupy  zonal wavenumbers
$|k_x|=0,1$  and  all  zonal wavenumbers $|k_x|\ge 2$ represent  phase incoherent motions, so that $P_K$ has
$K=1$.

The covariance of the stochastic excitation in~\eqref{eq:xi_cor} is assumed to be
spatially homogeneous, i.e.~\mbox{$Q(\xv_a,\xv_b)= Q(\xv_a-\xv_b)$},  and can be associated with
its Fourier power spectrum $\hat{Q}(\kv)$:
\be
Q(\xv_a-\xv_b) = \int \frac{\df^2 \mathbf{k}}{(2\pi)^2}\, \hat{Q}(\mathbf{k}) e^{i \mathbf{k}\bcdot (\xv_a-\xv_b)}\label{eq:Qhat}\ .
\ee
Unless otherwise indicated calculations are performed  with the 
anisotropic power spectrum:
\be
\hat{Q}(\kv) = \frac{ (4\pi/N_f) \, k_x e^{-k^2 d^2}}{ k_x/|k_x|-\textrm{erf}{(k_x d)}} \,\sum_{k_{f}\in K_{f}} \[\delta(k_x-k_{f})+\delta(k_x+k_{f})\bit\]\ ,\label{eq:spec_NIF}
\ee
with $k=|\kv|$, $d=0.2$, $K_{f}=\{2,3,\dots,14\}$ the zonal wavenumbers that are forced and $N_{f}$  the total number of  
excited zonal wavenumbers.  This spectrum is biased towards small $k_y$ numbers consistent with the assumption 
that the  forcing arises from baroclinic growth processes. The spatial excitation
covariance, $\hat{Q}(\kv)$,  has been normalized 
so that each $k_{f}$ injects equal energy and the total energy injection
rate is unity, i.e. $\hat{Q}(\mathbf{k})$ satisfies\footnote{A stochastic term $\sqrt{\e}\xi$ with spatial covariance given by~\eqref{eq:xi_cor} can be shown to inject average energy per unit area in the fluid at a rate $(L_xL_y)^{-1}\int\df^2\xv\;\<\psi\,\sqrt{\e}\xi \>= \e \[(2\pi)^{-2}\int \df^2 \mathbf{k}\;\hat{Q}(\mathbf{k})/(2k^2)\]$. Since dimensional $\xi$ has units $L^{-1} T^{-1/2}$ we obtain from~\eqref{eq:xi_cor} that  $Q$ has dimensions $L^{-2}$ therefore its Fourier transform $\hat{Q}$ is dimensionless. Hence~\eqref{eq:normaliza_Qhat} is valid for all values of the dimensional parameters.}:
\be
\int \frac{\df^2 \mathbf{k}}{(2\pi)^2}\frac{\hat{Q}(\mathbf{k})}{2k^2} =1\ .\label{eq:normaliza_Qhat}
\ee
With this normalization the rate of energy injection by the stochastic forcing  in~\eqref{eq:NLbarotropic},~\eqref{eq:s3ts},~\eqref{eq:PKNL},~\eqref{eq:PKQL} and~\eqref{eq:s3tf}  is $\e$ and is independent of the state of the system because $\xi$  has been assumed temporally delta-correlated.

We choose $\beta=10$, $r=0.15$ and $\nu=0.01$ as our parameters.
For $L=1200\,\textrm{km}$ and $T=6\,\textrm{d}$ these correspond to 
$\b=1.6\times10^{-11}\,\textrm{m}^{-1}\textrm{s}^{-1}$
and an \mbox{e-folding} time for linear damping  of  $40\,\textrm{d}$. The diffusion coefficient $\nu=0.01$ is chosen so that scales of the order of the grid are damped in one non-dimensional time and it corresponds to an e-folding time for scales of the order of $1000\,\textrm{km}$ (non-dimensional wavenumber $k_x=7$ in our channel) of approximately  $400\,\textrm{d}$.   
With these parameters the channel has zonal extent  about $7500\,\textrm{km}$, which corresponds to $1/4$ of the latitude circle
at $45^\circ$, one unit of velocity corresponds to $23\, \textrm{m}\,\textrm{s}^{-1}$ and non-dimensional  $\epsilon=1$ corresponds to an energy input rate of $1.03 \times 10^{-5}\,\textrm{W}\,\textrm{kg}^{-1}$. Simulations presented in this work are performed using a pseudospectral code with $N_x=N_y=64$
grid points. 

\section{S3T jet equilibria}

Fixed points of the S3T system correspond to statistical equilibria of
the barotropic dynamics. We study  these
statistical equilibria as a function of $\e$. For all values of $\e$ and
all homogeneous stochastic forcings  there  exist equilibria that are
homogeneous (both in $x$ and $y$) with 
\be
\Uv^{ h}=(0,0) \ ,\ \ \\ C^{ h}(\xv_a-\xv_b)=\e\int\frac{\df^2\kv}{(2\pi)^2} \frac{\hat{Q}(\kv) }{2(r+\nu k^2)}e^{i\kv\bcdot(\xv_a-\xv_b)}\ ,
\ee
where $\hat{Q}(\kv)$ is the power spectrum of the stochastic forcing,
defined in~\eqref{eq:Qhat}.

\begin{figure}
\centering
\includegraphics[width=.45\textwidth]{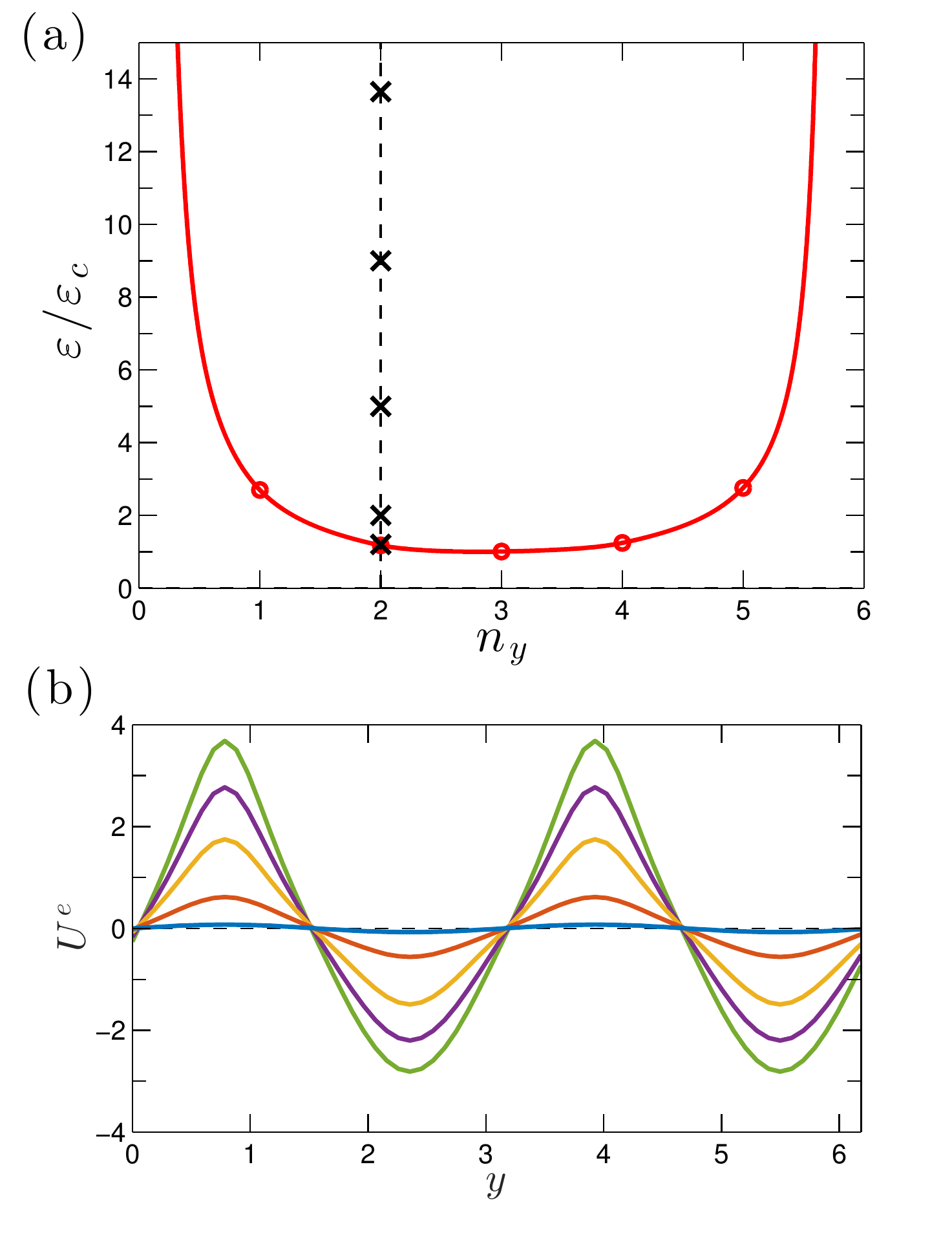}
\vspace{-1em}
\caption{(a) Normalized turbulent energy input rates, $\e/{\ecz}$, at which the homogeneous state becomes unstable to jet ($n_x=0$) perturbations as a function of the jet meridional wavenumber $n_y$. Dots indicate wavenumbers allowed in the model, ${\ecz}$ is the minimum energy input rate for jet emergence. Jets first emerge in an unrestricted eigencalculation at $\ecz=0.2075$ with unallowed wavenumber $n_y=2.82$. For $\e/{\ecz}<1.18$ the homogeneous state is stable to $n_y=2$ mean flow perturbations and $n_y=2$ jet equilibria do not exist. (b) The $n_y=2$ zonal jet S3T equilibrium structure at $\e/{\ecz} = 1.2,2,5,9,13.65$ (marked with $\times$ in panel (a)). Increasing supercriticality results in increasing equilibrium jet amplitude  and deviation of the jet structure from the sinusoidal eigenmode form.}\label{fig:ecz_nu2_0p01_Ue}
\end{figure}

However, these equilibria become unstable when $\e$ exceeds a critical
value. For values of $\e$ exceeding this critical value 
zonal jets arise from a supercritical bifurcation~\citep{Farrell-Ioannou-2003-structural,Farrell-Ioannou-2007-structure,Srinivasan-Young-2012,
Parker-Krommes-2013,Parker-Krommes-2014-generation,Constantinou-etal-2014}.
These jets are constrained by the periodic domain of our simulations to take discrete values of meridional wavenumber, $n_y$. 
The critical curve in the $(\e,n_y)$~plane separating the region
in which only stable homogeneous turbulence equilibria exist from the region in
which stable or unstable jet equilibria exist is shown for the
chosen parameters in Fig.~\ref{fig:ecz_nu2_0p01_Ue}. This marginal curve
was calculated using the eigenvalue relation for inhomogeneous
perturbations to the homogeneous S3T equilibrium in the presence of
diffusive dissipation, in the manner of~\citet{Srinivasan-Young-2012}
and~\citet{Bakas-Ioannou-2014-jfm} with the wavenumber $n_y$ taking
continuous values, but with the understanding that only integer values
of $n_y$ satisfy the quantization conditions of the channel. S3T
instability  of the homogeneous state first occurs at $n_{y} = 2.82$
for $\ecz =  0.2075$,  which corresponds to $2.15 \times 10^{-6}\,\textrm{W}\,\textrm{kg}^{-1}$.  Jets with $n_y=3$ emerge at $1.005\ecz$ and  jets
with $n_y=2$   at $1.18\ecz$. Examples of $n_y=2$ jet
equilibria are shown in Fig.~\ref{fig:ecz_nu2_0p01_Ue}b. The $n_y=2$ jet
equilibria have  mean flows and covariances that are periodic in $y$
with period $\alpha=\pi$ and satisfy the time-independent S3T
equations:\begin{subequations}
\label{eq:jet_eq}\begin{align}
&\frac1{2}\[\(\partial_{x_a}\Delta_a^{-1}+\partial_{x_b}\Delta_b^{-1}\)C^e\]_{\xv_a=\xv_b}=  r  U^e - \nu \partial^2_{y} U^e  \ ,\\
&\(A_a^e+A_b^e\bit\)C^e= - \e\,Q\ ,
\end{align}\end{subequations}
with
\begin{align}
A^e=-U^e\partial_{x} - \[\bit\beta - (\partial^2_{y} U^e) \]\partial_{x}\Del^{-1}-r + \nu \Del\ .\ \label{eq:Ajet}
\end{align}
A basic property of the jet equilibria, which is shared by all  S3T
equilibria, is that they are hydrodynamically stable 
(cf.~\cite{Farrell-Ioannou-2015-book}). Stability is enforced at the
discrete wavenumbers consistent with the finite domain of the problem and not
necessarily on the continuum of wavenumbers appropriate for an unbounded domain.

\section{S3T stability of  the jet equilibria}

We are interested in the S3T stability of these $n_y=2$ jet equilibria
to non-zonal perturbations. The stability of jet equilibria to
homogeneous in $x$ perturbations has been investigated previously
by~\cite{Farrell-Ioannou-2003-structural,Farrell-Ioannou-2007-structure}
for periodic domains and by~\citet{Parker-Krommes-2014-generation,Parker-2014} 
for infinite domains. A comprehensive methodology for determining the stability of
jet equilibria to zonal and non-zonal perturbations was developed by~\citet{Constantinou-2015-phd}. Recalling these results, perturbations
$(\d Z,\d C)$ about the  equilibrium state $(U^e,C^e)$, satisfying
equations~\eqref{eq:jet_eq}, evolve according to\footnote{These perturbations equations are  valid  for equilibria  inhomogeneous  in
both $x$ and $y$ directions.  In the case of jet  equilibria the projection operators are redundant.}:\ifdraft
\begin{subequations}\label{eq:s3t_dZdCgen}
\begin{align}
\partial_t \,\d Z & =P_K \[ \bit A^e\,\d Z + \Rcal( \d C ) \]\ ,\label{eq:s3t_pert_dZgen}\\
\partial_t \,\d C & = (I-P_{K a}) \(   A^e_a  \d C +  \d A_a C^e  \) + (I-P_{K b}) \( A^e_b \d C + \d A_b C^e \)  \ ,\label{eq:s3t_pert_dCgen}
\end{align}\end{subequations}
\else
\begin{subequations}\label{eq:s3t_dZdCgen}
\begin{align}
\partial_t \,\d Z & =P_K \[ \bit A^e\,\d Z + \Rcal( \d C ) \]\ ,\label{eq:s3t_pert_dZgen}\\
\partial_t \,\d C & = (I-P_{K a}) \(   A^e_a  \d C +  \d A_a C^e  \)   \nonumber\\
&\hspace{4em}+ (I-P_{K b}) \( A^e_b \d C + \d A_b C^e \)  \ ,\label{eq:s3t_pert_dCgen}
\end{align}\end{subequations}\fi
with $\Rcal$ as in~\eqref{eq:uv}), $A^e$ defined in~\eqref{eq:Ajet} and  
\begin{align}
\d A\equiv-\d \Uv\bcdot\nablav+ (\Delta \d\Uv)\bcdot\nablav \Del^{-1}\ ,\ \label{eq:dAjet}
\end{align}
where $\d \Uv= \zhat\times\nablav\Del^{-1}\d Z$ 
 is the perturbation velocity field.

\begin{figure*}
\centerline{\includegraphics[width=0.55\textwidth]{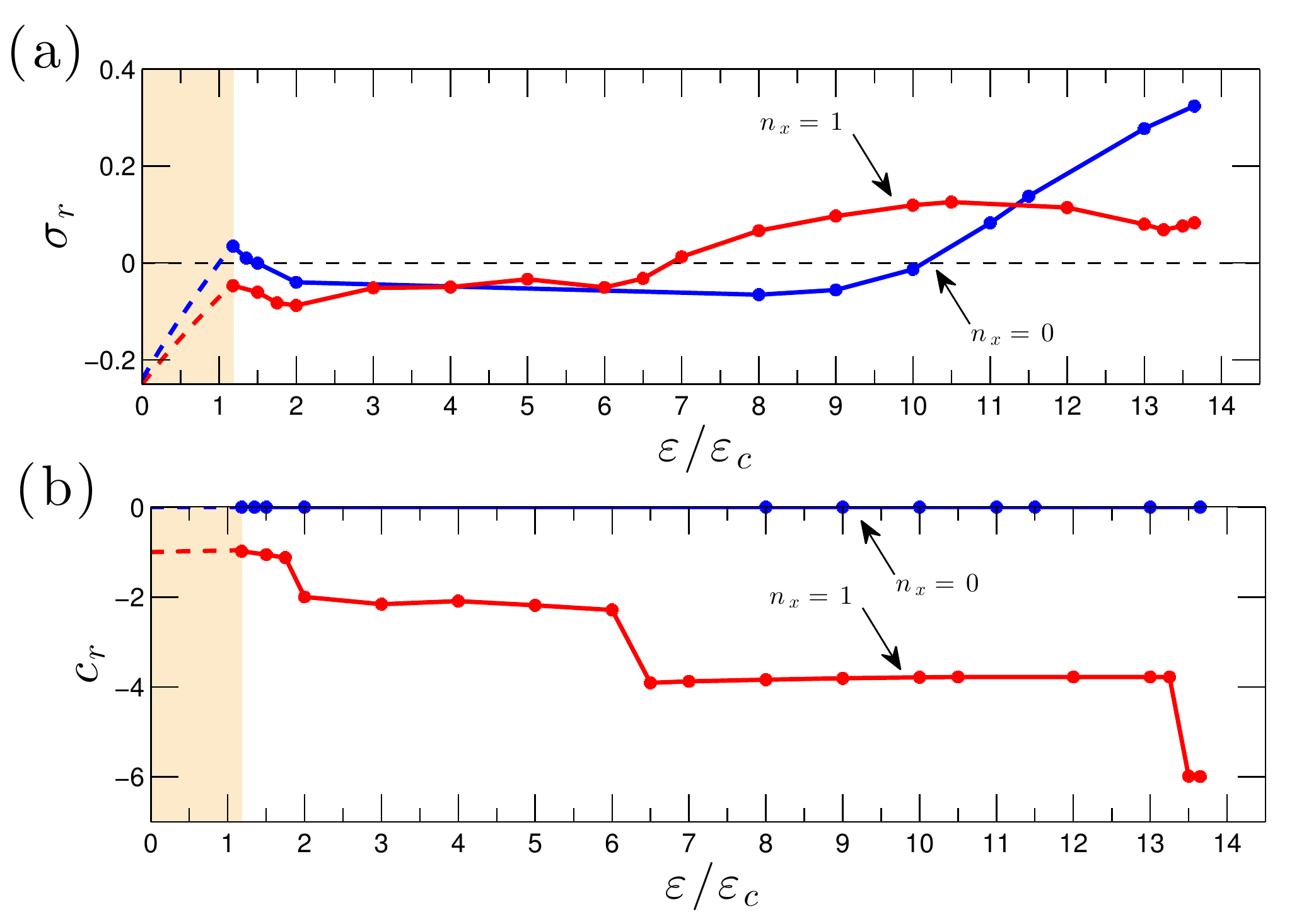}}
\vspace{-5mm}
\caption{(a) Maximum S3T growth rates $\s_r$ as a function of $\e/\ecz$ for $n_x=0$ and $n_x=1$ perturbations to the $n_y=2$ equilibrium jets. The  jet is unstable to jet ($n_x=0$) perturbations for $1.18\le \e/\ecz \le 1.44$  and  $\e/\ecz\ge10.14$ and to $n_x=1$ wave perturbations for $\e/\ecz\ge6.80$. (b) The corresponding phase speeds $c_r$ of the most unstable S3T eigenfunction.} \label{fig:S3Tgrowth}
\vspace{-1em}

\centerline{\includegraphics[width=18pc,trim = 0mm 0mm 2mm 0mm, clip]{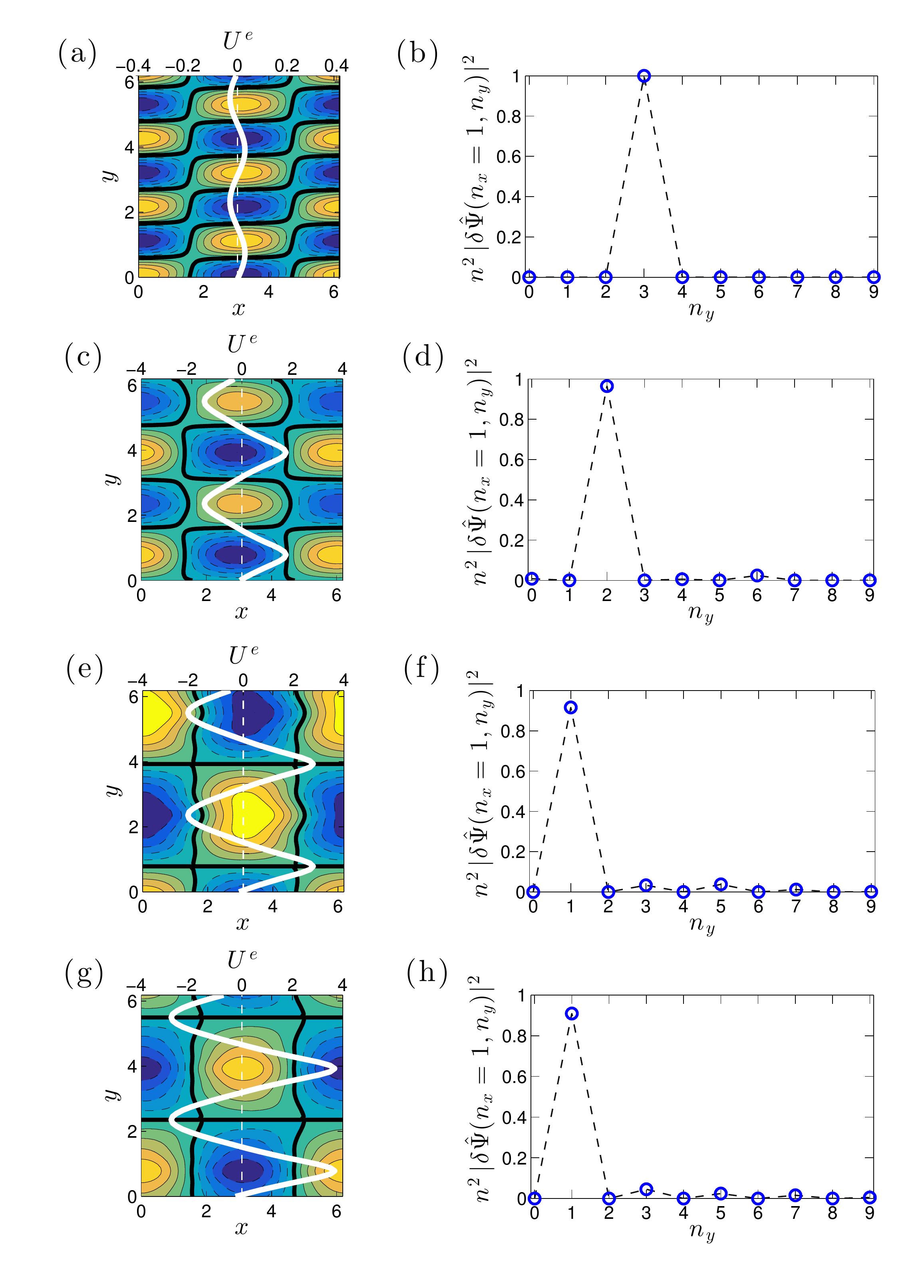}}
\vspace{-1em}
\caption{(a):~Contour plot  of the streamfunction of the least stable non-zonal $n_x=1$ S3T mean flow wave eigenfunction of the $n_y=2$ jet equilibrium at $\e/\ecz=1.2$. This wave has growth rate $\s_r=-0.047$ and phase speed $c_r=-0.98$. The equilibrium jet is  shown in solid white. Positive (negative) contours are indicated with solid (dashed) lines and the zero contour with a thick solid line. (b):~The power spectrum of the mean flow eigenfunction. This jet equilibrium is unstable to $n_x=0$ perturbations but stable to  $n_x=1$ perturbations. The least stable $n_x=1$ eigenfunction is Bloch $q_y=1$ with power at $n_y=3$. (c,d):~Same for the least stable $n_x=1$ S3T eigenfunction of the equilibrium at $\e/\ecz=5$. The jet is stable both to $n_x=0$ and $n_x=1$ perturbations and the least stable $n_x=1$ eigenfunction ($\s_r=-0.033$, $c_r=-2.18$) is Bloch $q_y=0$ with power at $n_y=2$. (e,f):~Same for the maximally growing $n_x=1$ S3T eigenfunction of the jet at $\e/\ecz=9$. The jet is stable to $n_x=0$ perturbations but unstable to $n_x=1$ perturbations and the most unstable $n_x=1$ eigenfunction ($\s_r=0.099$, $c_r=-3.81$) is Bloch $q_y=1$ with power at $n_y=1$. (g,h):~Same for the maximally growing $n_x=1$ S3T eigenfunction of a strong equilibrium jet at $\e/\ecz=13.65$. The most unstable $n_x=1$ eigenfunction ($\s_r=0.083$, $c_r=-5.99$) is  Bloch $q_y=1$ with power at $n_y=1$. In this case, $n_x=0$ perturbations are more unstable with $\s_r=0.324$.}\label{fig:eig_nx1_spec}
\end{figure*}

Because of the homogeneity of  the jet equilibria in the zonal, $x$,
direction  the mean flow eigenfunctions are harmonic functions in $x$
and also because the equilibrium mean flow and covariance are periodic
in $y$ with period $\alpha$, i.e. $U^e(y+\a)=U^e(y)$, Bloch's theorem
requires that each eigenfunction is a plane wave in $y$, $e^{i q_y y}$,
modulated by a periodic function with period $\alpha$ in
$y$~\citep{Cross-Greenside-2009}. Therefore, the eigenfunctions  take the
form:\ifdraft\begin{subequations}
\begin{align}
\d Z &= e^{i n_x x + i q_y y +\sigma t} \dZ_{n_x,q_y}(y) \ ,\label{eq:S3T_dZ}\\
\d C&= e^{i n_x(x_a+x_b)/2 + i q_y (y_a+y_b)/2 + \s t} \[\bit\dC_{n_x,q_y}(x_a-x_b,y_a,y_b) + \dC_{n_x,q_y}(x_b-x_a,y_b,y_a)\]  \ ,
\end{align}\label{eq:S3Teigen_nx}\end{subequations}
\else
\begin{subequations}
\begin{align}
\d Z &= e^{i n_x x + i q_y y +\sigma t} \dZ_{n_x,q_y}(y) \ ,\label{eq:S3T_dZ}\\
\d C&= e^{i n_x(x_a+x_b)/2 + i q_y (y_a+y_b)/2 + \s t} \vphantom{\[\bit\dC_{n_x,q_y}(x_a-x_b,y_a,y_b) + \dC_{n_x,q_y}(x_b-x_a,y_b,y_a)\]}\nonumber\\
& \vphantom{e^{i n_x(x_a+x_b)/2 + i q_y (y_a+y_b)/2 + \s t}} \[\bit\dC_{n_x,q_y}(x_a-x_b,y_a,y_b) + \dC_{n_x,q_y}(x_b-x_a,y_b,y_a)\]  \ ,
\end{align}\label{eq:S3Teigen_nx}\end{subequations}
\fi
 with $|n_x| \le K$, $\dZ_{n_x,q_y}(y)$  periodic in $y$ with period $\alpha$ and $\dC_{n_x,q_y}(x_a-x_b,y_a,y_b)$   periodic in $y_a$
 and $y_b$ with period $\alpha$. We have chosen $\d C$ to be a symmetric function of $\xv$ under the 
exchange $\xv_a\leftrightarrow\xv_b$\footnote{The covariance eigenfunction does not need to be symmetric or Hermitian
in its matrix representation, but both symmetric and asymmetric parts
have the same growth rate. For a discussion of the properties of
covariance eigenvalue problems cf.~\cite{Farrell-Ioannou-2002-perturbation-II}.}.
The zonal wavenumber, $n_x$,  takes integer values in order
to satisfy the periodic boundary conditions  in $x$, and the Bloch
wavenumber, $q_y$,   takes integer values in the interval  $|q_y| \le
\pi/\alpha$, in order to satisfy the periodic boundary conditions in $y$
\citep{Constantinou-2015-phd}. The  eigenvalue $\s$ determines the S3T stability of the jet as a function of  
$n_x$  and $q_y$. 
The jet is unstable when $\sigma_r
\equiv\real(\s) >0$ and the S3T eigenfunction propagates in $x$ with
phase velocity $c_r\equiv-\imag(\s)/n_x$ for $n_x\ne 0$. When $n_x=0$
the eigenfunctions are  homogeneous in the zonal, $x$, direction and
correspond to a perturbation zonal jet. When $n_x\ne 0$ the
eigenfunctions are inhomogeneous in both $x$ and $y$ and correspond to a
 wave. These perturbations, when unstable, can form non-zonal
large-scale structures that coexist with the mean flow, as 
in \cite{Bakas-Ioannou-2014-jfm}. For jets with meridional periodicity $\a=\pi$,  $q_y$
can take only the values  $q_y=0,1$ and  because  these jets have a
Fourier spectrum with  power only at the even wavenumbers, a $q_y=0$
Bloch eigenfunction has power only at even wavenumbers, while a $q_y=1$
Bloch eigenfunction has power only at odd wavenumbers.  

The maximum growth rate, $\sigma_r$, of the S3T eigenfunction
perturbations to the S3T equilibrium jet with $n_y=2$ (cf.~Fig.~\ref{fig:ecz_nu2_0p01_Ue}b) 
is plotted in Fig.~\ref{fig:S3Tgrowth}a as a function of supercriticality $\e/\ecz$
for both perturbations of jet form ($n_x=0$) and non-zonal form (with
$n_x=1$). Consider first the stability of the S3T jet to jet
perturbation, that is to $n_x=0$ eigenfunctions.  Recall that the jets
with $n_y=2$ emerge at $\e/\ecz=1.18$, and for $\e/\ecz < 1.18$ (shaded
region in  Fig.~\ref{fig:S3Tgrowth}a)  there are no $n_y=2$ equilibria.
The dashed line shows the smallest decay/fastest growth  rate of perturbations to the
homogeneous equilibrium state that exists prior to jet formation. The
most unstable eigenfunctions of the homogeneous equilibria at these $\e$
are jets with wavenumber  $n_y=3$ (not shown, cf.~Fig.~\ref{fig:ecz_nu2_0p01_Ue}a). 
The small amplitude equilibrated
$n_y=2$ jets that form when $\e$ marginally exceeds the critical
$\e/\ecz =1.18$ are  unstable to  jet formation 
at wavenumber $n_y=3$ with jet eigenfunction similar to
the maximally growing $n_y=3$ eigenfunction  of the homogeneous
equilibrium. This S3T instability of the small amplitude $n_y=2$  jet equilibria
to $n_y=3$  jet eigenfunctions, which is induced by the $n_y=3$ instability of the nearby
homogeneous equilibrium, was identified by
\cite{Parker-Krommes-2014-generation} as the universal Eckhaus
instability of the equilibria that form near a supercritical
bifurcation. The Eckhaus unstable S3T $n_y=2$ jets are attracted to
the S3T $n_y=3$ stable jet equilibrium over the small interval $1.18<\e/\ecz < 1.44$.  
At higher supercriticalities in the interval 
$1.44<\e/\ecz < 10.14$ the  $n_y=2$ jets become stable\footnote{The periodic boundary
conditions  always allow the existence of a jet  eigenfunction with zero
growth and with structure that of the $y$ derivative of the equilibrium
jet and covariance. This eigenfunction leads to a translation of the
equilibrium jet and its associated covariance in the $y$ direction. This
existence of this neutral eigenfunction can be a verified by taking the
$y$ derivative of~\eqref{eq:jet_eq}. We do not include this obvious
neutral eigenfunction in the stability analysis.} to $n_x=0$ eigenfunctions. The jets eventually become
unstable to $n_x=0$ eigenfunctions for $ \e/\ecz > 10.14$. The most
unstable $n_x=0$ eigenfunction at $\e/\ecz=11$ is a Bloch $q_y=1$
eigenfunction, dominated by a $n_y=1$ jet that will make the jets of the
$n_y=2$ equilibrium  merge to form a $n_y=1$ jet equilibrium
(cf.~\cite{Farrell-Ioannou-2007-structure}).

The maximum growth rate of the jet equilibria to $n_x=1$ non-zonal
eigenfunctions is also shown  in Fig.~\ref{fig:S3Tgrowth}a. Unlike the
jet eigenfunctions, which are stationary  with respect to the mean flow, 
these eigenfunctions propagate retrograde with respect to the jet minimum; the phase  velocity of the  eigenfunction with maximum real part eigenvalue is plotted as a function of $\e/\ecz$ in Fig.~\ref{fig:S3Tgrowth}b.
Eigenfunctions with $n_x=1$ are stable for jets with $\e/\ecz \le 6.80$
and when  they become unstable  the jet is still stable to jet ($n_x=0$)
perturbations. The structure of the least damped/fastest growing eigenfunctions at
various $\e/\ecz$ are shown in  Fig.~\ref{fig:eig_nx1_spec}. In
Fig.~\ref{fig:eig_nx1_spec}a,b is shown the least stable eigenfunction
of  the weak jet at $\e/\ecz=1.2$. The eigenfunction is Bloch $q_y=1$
with  almost all power at $n_y=3$. The phase velocity of this 
eigenfunction is $c_r=-0.98$ which is slightly slower than the Rossby
phase speed  $-1$  (i.e., $-\b/k^2$, with $\beta=10$, $k_x=1$, $k_y=3$).
In Fig.~\ref{fig:eig_nx1_spec}c,d is shown the least stable $n_x=1$ mode
for the jet at $\e/\ecz=5$ which is Bloch $q_y=0$ with almost all power
at $n_y=2$ and phase speed $c_r=-2.18$, which corresponds to a slightly 
modified  Rossby  phase speed with effective PV gradient of  $\b_{\rm eff}=10.9$ 
instead of the $\b=10$ of the uniform flow. 
In Fig.~\ref{fig:eig_nx1_spec}e-h are 
shown the maximally unstable $n_x=1$ eigenfunctions for the jets at
$\e/\ecz=9$ and $\e/\ecz=13.65$. Both are Bloch $q_y=1$ with almost all
power at $n_y=1$. At $\e/\ecz=9$ the mode is trapped in the retrograde
jet, a region of reduced PV gradient, and the structure of this mode as
well as its phase speed corresponds, as shown in the next section, to
that of  an external Rossby wave  confined in this equilibrium flow. At
$\e/\ecz=13.65$ the eigenfunction is trapped in the prograde jet, a
region of high PV gradient, and the structure of this mode as well as
its phase speed corresponds to that of  an external Rossby wave in this
equilibrium flow.

\section{The mechanism destabilizing  S3T jets  to $n_x= 1$ non-zonal  perturbations \label{sec:s3tstab_nxne0}}

We now examine the stability properties of the $n_y=2$ equilibrium jet
maintained in S3T at $\e/\ecz=9$. At  $\e/\ecz=9$ the jet is stable to
$n_x=0$ jet S3T perturbations  but unstable to $n_x=1$ non-zonal
perturbations with  maximally growing eigenfunction  growth rate 
$\s_r=0.099$ and phase speed $c_r=-3.806$, which is retrograde at speed $1.61$ 
with respect to the minimum velocity of the jet. 

\begin{figure*}
\centering
\includegraphics[width=.67\textwidth]{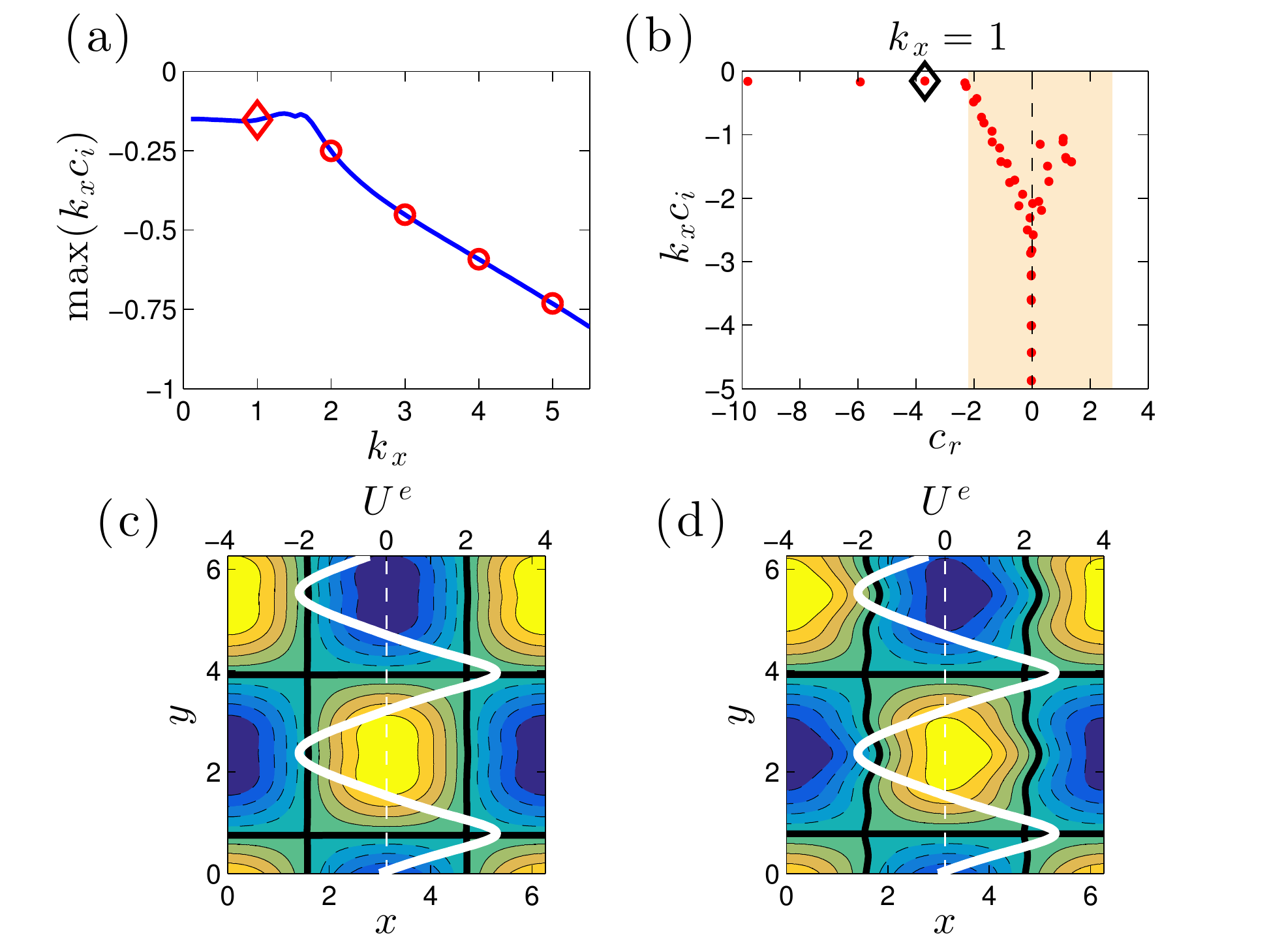}
\vspace{-1em}
\caption{(a) The hydrodynamic stability of $U^e$ at $\e/\ecz=9$. Shown are the maximal modal growth rates, $k_x c_i$, of operator $A^e$ as a function of $k_x$. Circles indicate the growth rate at the $k_x$ retained in the perturbation dynamics; diamond indicates the growth rate at $k_x=1$. The equilibrium jet is hydrodynamically stable but S3T unstable to $n_x=1$ perturbation. (b) The growth rates, $k_x c_i$, and phase speeds, $c_r$, of the least damped eigenvalues of $A^e$ for $k_x=1$ perturbations. The shaded area indicates the region $\min{(U^e)}\le c_r \le\max{(U^e)}$. The jet $U^e$ is shown in white in both panels (c) and (d). 
The streamfunction of the maximally growing S3T $n_x=1$ eigenfunction is shown in (d). This S3T eigenfunction arises from destabilization of the least damped mode of $A^e$ with $k_x c_i =-0.15$ and $c_r=-3.70$,  indicated with the diamond in (b)  and shown in (c). The $n_x=1$ S3T instability with $\s_r=0.099$ and phase speed $c_r=-3.81$  is supported in this case solely by energy transfer from the mean flow $U^e$ (at the rate $\s_{10}=0.303$) 
against the negative energy transfer from the small scale perturbation field (at the rate $\s_{1>}=-0.016$) and dissipation (at the rate $\s_{1\textrm{D}}=-0.188$), with the growth rate of the S3T instability 
being $\s_{r} = \s_{10}+\s_{1>}+\s_{1\textrm{D}}$.}
\label{fig:spec_comb_9p00} 
 \centering
\centering
\includegraphics[width=.67\textwidth]{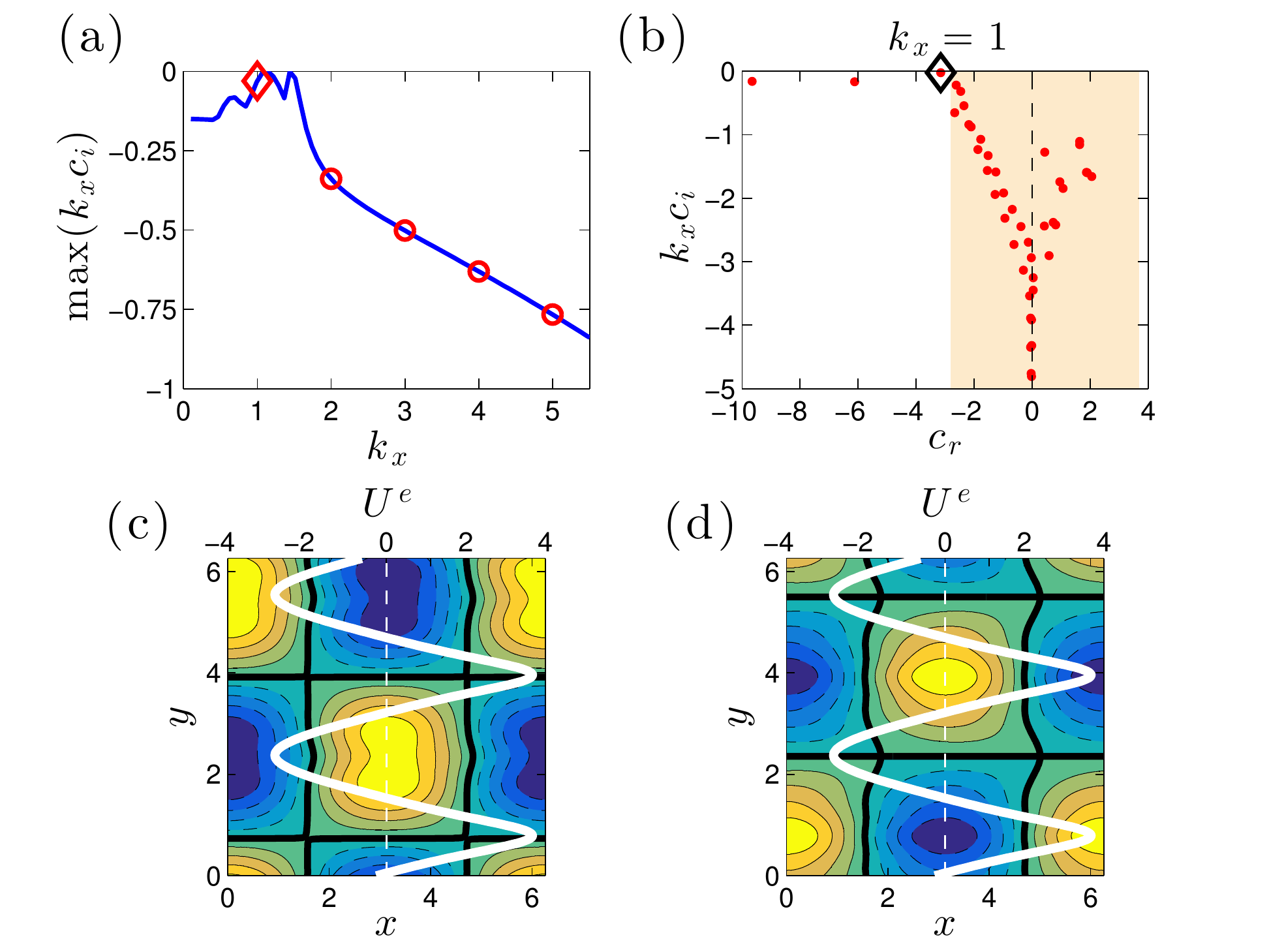}
\vspace{-1em}
\caption{(a) The hydrodynamic stability of  $U^e$ at $\e/\ecz=13.65$. Shown are the maximal modal growth rates, $k_x c_i$, of operator $A^e$ as a function of $k_x$. Circles indicate the growth rate at the $k_x$ retained in the perturbation dynamics; diamond indicates the growth rate at $k_x=1$. The equilibrium jet is hydrodynamically stable but S3T unstable to both $n_x=0$ and $n_x=1$ perturbations. (b) The growth rates, $k_x c_i$, and phase speeds, $c_r$, of the least damped eigenvalues of $A^e$ for $k_x=1$ perturbations. The shaded area indicates the region $\min{(U^e)}\le c_r \le\max{(U^e)}$.  The jet $U^e$ is shown in white in both panels (c) and (d). The streamfunction of the maximally growing S3T $n_x=1$ eigenfunction is shown in (d). This S3T eigenfunction  arises from destabilization of the second least damped mode of $A^e$ with $k_x c_i = -0.165$ and $c_r=-6.12$,  indicated with a diamond in (b) and shown in (c). The $n_x=1$ S3T instability with $\s_r=0.083$ and $c_r=-5.99$ is supported in this case by both energy transfer from the mean flow $U^e$ (at the rate $\s_{10}=0.160$) and energy transfer from the small scale perturbation field (at the rate $\s_{1>}=0.115$). The dissipation rate is $\s_{1\textrm{D}}=-0.192$.}
\label{fig:spec_comb_13p65}
\end{figure*}

Because the jet $U^e$ is an S3T equilibrium the operator $A^e$ is
necessarily stable to perturbation at zonal wavenumbers that are
retained in the perturbation dynamics, i.e. $k_x\in K_{f}$. The maximum
growth rate of operator $A^e$ as a function of $k_x$ for the  jet
$\e/\ecz=9$ is shown in Fig.~\ref{fig:spec_comb_9p00}a, with the integer
valued wavenumbers that are included  in the S3T dynamics and are
responsible for the stabilization of the jet indicated with a circle in
this figure. This equilibrium jet, despite its robust hydrodynamic
stability at all wavenumbers, in both the mean and perturbation
equations, and especially its hydrodynamic stability to $n_x=1$
perturbations, is nevertheless S3T unstable at $n_x=1$. 

Although it is not formed as a result of a traditional hydrodynamic instability,  this S3T instability is very close in structure to the least stable eigenfunction of $A^e$  at $n_x=1$, as it
can be seen in  Fig.~\ref{fig:spec_comb_9p00}c,d. The spectrum of $A^e$
at $n_x=1$ is shown in 
Fig.~\ref{fig:spec_comb_9p00}b.  The eigenfunctions associated with this spectrum 
consist of viscous shear modes with
phase speeds within the flow and a discrete number of external Rossby
waves with phase speeds retrograde with respect to the minimum
velocity of the flow (cf.~\cite{Kasahara-1980}). In this case there are
exactly 5 external Rossby waves with phase speeds
$c_r=-3.70,-9.80,-5.92,-2.33,-2.37$ all decaying with
$k_x c_i=-0.15,-0.16,-0.17,-0.18,-0.24$ respectively. We identify the S3T
$n_x=1$ unstable eigenfunction, shown in Fig.~\ref{fig:spec_comb_9p00}d, 
which has phase speed $c_r=-3.81$ with S3T destabilization of the 
least stable of the external Rossby waves, shown in Fig.~\ref{fig:spec_comb_9p00}c, that has phase speed $c_r=-3.70$. This instability arises 
by Reynolds stress feedback 
that exploits the least damped mode of $A^e$, 
which is already extracting some energy from the jet through the hydrodynamic instability process, thereby making it S3T unstable. This
feedback process transforms a mode of the system that while extracting 
energy from the mean nevertheless was decaying at a rate $k_x   c_i = -0.15$ into
an unstable mode growing at rate $\sigma_r=0.099$. Consistently, note in
Fig.~\ref{fig:spec_comb_9p00}d  that the streamfunction  of the S3T
eigenfunction is tilting against the shear indicative of its gaining
energy from the mean flow.

We quantify the energetics of the S3T instability in order to examine the instability mechanism in more detail. The contribution to the growth rate of this $n_x=1$ eigenfunction from interaction with the mean equilibrium jet is
\be
\sigma_{10} = \frac{1}{2}\frac{\( \bit A_{\textrm{inv}}(U^e)\,\d Z,\d Z\) + \(\bit \d Z,A_{\textrm{inv}}(U^e)\d Z\) }{\(\bit \d Z,\d Z\)}\ , \label{eq:s1a}
\ee
where $\( f\, ,\, g\) \equiv- (2 \pi)^{-2} \int\df^2\xv \;\frac1{2} f\,\Del^{-1} g$ is the inner product in energy metric and
\begin{align}
A_{\textrm{inv}}(U)=-U\partial_{x} - \[\bit\beta - (\partial^2_{y} U) \]\partial_{x}\Del^{-1} \ , \label{eq:Ainv}
\end{align} 
is the inviscid part of~\eqref{eq:op_A} with $V=0$. The
contribution to the growth rate of the $n_x=1$ eigenfunction from 
Reynolds stress mediated interaction with the small scales is  
\be
\sigma_{1>} = \frac{1}{2}\frac{\(\bit \d Z, R(\d C)\)+ \(\bit  R(\d C), 
\d Z\)}{\(\bit \d Z,\d Z\)}\ .\label{eq:s1r}
\ee
The net growth rate of  the perturbation  $n_x=1$ eigenfunction is then 
$\s_r = \s_{10}+ \s_{1>}+\s_{1\textrm{D}}$, with
\be
\sigma_{1\textrm{D}} =\frac{1}{2}\frac{\( \bit A_{\textrm{D}}\,\d Z,\d Z\) + \(\bit \d Z,A_{\textrm{D}}\d Z\) }{\(\bit \d Z,\d Z\)}~,\label{eq:s1d}
\ee
the loss to dissipation, where
\begin{align}
A_{\textrm{D}}=-r + \nu \Del\ ,\ \label{eq:Adis}
\end{align}
is the dissipation part of operator~\eqref{eq:op_A}. 

For the S3T unstable eigenfunction shown in
Fig.~\ref{fig:spec_comb_9p00}d, the growth rate $\s_r=0.099$ arises
solely from interaction with the mean flow, which contributes $\s_{10}=0.303$, while  the energy transfer from the small scale perturbation field  contributes negatively, $\s_{1>}=-0.016$, with dissipation accounting for the 
remainder $\s_{1\textrm{D}}=-0.188$.
Interestingly, this S3T unstable mode is solely supported in its
energetics by induced non-normal interaction with the mean jet and loses
energy to the Reynolds stress feedback which is responsible for the instability. 
This remarkable mechanism arises from  eddy flux interaction transforming 
damped waves  into exponentially growing waves 
by changing the wave structure so as to tap the energy of the mean jet.  This novel mechanism destabilizes the wave
even though the  direct effect of the
Reynolds stresses  is to stabilize it. This mechanism of
destabilization differs   from that acting in  more familiar S3T
instabilities in which jets and waves arise directly  from their interaction with the  incoherent eddy field.

This same mechanism is responsible for the S3T destabilization of the
$n_x=1$  perturbation to  the  jet equilibrium at $\e/\ecz=13.65$.
However, at  $\e/\ecz=13.65$ the jet is unstable to both $n_x=0$ (with
maximum growth rate $\s_r= 0.324$) and  to $n_x=1$ non-zonal
perturbations (with maximum growth rate  $\s_r=0.083$ and phase speed
$c_r= -5.99$, which is retrograde by $3.18$ with respect to the minimum
velocity of the jet). This equilibrium flow is also hydrodynamically
stable at all the zonal wavenumbers allowed by periodicity
(cf.~Fig.~\ref{fig:spec_comb_13p65}a). This $n_x=1$ unstable
eigenfunction  (cf.~Fig.~\ref{fig:spec_comb_13p65}d) arises from
destabilization   of the second least damped mode, which is the damped
external Rossby mode indicated in Fig.~\ref{fig:spec_comb_13p65}b and
shown in Fig.~\ref{fig:spec_comb_13p65}c. The energetics of the
instability indicate that the growth of this $n_x=1$ structure arises
almost equally from energy transferred from the mean equilibrium jet to
the $n_x=1$ perturbation ($\s_{10}=0.160$) and energy transferred by the small scales ($\s_{1>}=0.115$) while dissipation accounts for the 
remainder $\s_{1\textrm{D}}=-0.192$.

\section{Equilibration of  the S3T instabilities of the equilibrium jet\label{sec:equil_nx1}}

We  next  examine  equilibration of the  $n_x=1$ S3T instability at
$\e/\ecz=9$ and the equilibration of the S3T instabilities at
$\e/\ecz=13.65$, which has  both  $n_x=0$ and $n_x=1$ unstable eigenfunctions.

Consider the energetics of these large scales consisting of the $k_x=0$
and $k_x=1$ Fourier components. Denote the $k_x=0$ and $k_x=1$ 
components of vorticity of~\eqref{eq:s3tmf} as $Z^0$ and $Z^1$ and the
corresponding vorticity flux divergence of the incoherent components as $R^0$,
$R^1$  and with $Z^e \equiv-\partial_y U^e$ the vorticity of the equilibrium zonal
jet. The energetics of the equilibration of the S3T instabilities is examined by first
removing the constant flux to the large scales from the small scales
that maintains  the equilibrium flow $U^e$. For that reason the vorticity flux divergence associated with
 the  deviation of the instantaneous covariance from
$C^e$ will be considered in the equilibration process.

Consider  first the energetics of the $k_x=1$ component of the large-scale
flow. The first contribution to the energy growth of this component is the
energy transferred from the $k_x=0$ component of the flow. This occurs at
rate: 
\be
{\cal E}_{10}= \( A_{\textrm{inv}}(U^0) Z^{1}, Z^{1}\) + \( Z^{1}, { A_{\textrm{inv}}(U^0)}  Z^1\)\  ,\label{eq:e1a}
\ee
with  $A_{\textrm{inv}}$ defined in~\eqref{eq:Ainv} and $U^0$ the total
$k_x=0$ component of the zonal velocity. The second energy source is energy
transferred to $k_x=1$ from the small scales (i.e. those with $|k_x| >K$),  which occurs at  rate:
\be
{\cal E}_{1>}=\( Z^{1}  ,R^1\) + \(R^{ 1} , Z^1\)\ ,\label{eq:e1r}
\ee
with $R^1\equiv R^1(C-C^e)$  the vorticity flux divergence  produced by
covariance $C-C^e$. Finally, energy is dissipated at the rate:
\be
{\cal E}_{1\textrm{D}} = \(A_{\textrm{D}} Z^{1},Z^1\) + \(Z^1,A_{\textrm{D}} Z^1\)~,\label{eq:e1d}
\ee
with $A_{\textrm{D}}$ defined in~\eqref{eq:Adis}. 

The energy flowing  to the $k_x=0$ component  consist first of ${\cal E}_{01}$,  the energy transfer rate to this component from the $k_x=1$ component, which is equal to $-{\cal E}_{10}$ (being equal and opposite to the energy transfer rate  to  $k_x=1$ from the $k_x=0$ component), and second of the energy transferred to $k_x=0$ by 
the
small scales,  with contribution to the growth  rate:
\be
{\cal E}_{0>}= \( Z^{0}, R^0\) + \( R^{ 0} ,Z^0\)\ ,\label{eq:e0r}
\ee
with $R^0\equiv R^0(C-C^e)$. 
Having removed the energy source sustaining the
equilibrium flow, the energy of $Z^0$ is dissipated at  rate:
\be
{\cal E}_{0{\textrm{D}}} = \(A_{\textrm{D}}(Z^0-Z^e) , Z^0 \) + \( Z^{0}, A_{\textrm{D}} ( Z^0-Z^e)\)~.\label{eq:e0d}
\ee

The instantaneous rate of change of the energy  of the  $Z^0$ and  $Z^1$
components are then $\df E_0/\df t= {\cal E}_{01}+{\cal E}_{0>}+{\cal E}_{0{\textrm{D}}}$ 
and $\df {E}_1/\df t = {\cal E}_{10}+{\cal E}_{1>}+{\cal E}_{1{\textrm{D}}}$. By
dividing each term of $\df {E_1}/\df t$ with $2 (Z^1,Z^1)$ we obtain, corresponding
to~\eqref{eq:s1a},~\eqref{eq:s1r},~\eqref{eq:s1d}, the instantaneous growth
rates $\s_{10}$, $\s_{1>}$ and $\s_{1{\textrm{D}}}$ and by dividing
$\df E_0/\df t$ with $2(Z^0-Z^e,Z^0-Z^e)$ the growth rates $\s_{01}$, $\s_{0>}$ and 
$\s_{0{\textrm{D}}}$. As equilibration is approached
the sum of these growth rates approaches zero, while the evolution of
the growth rates indicates the role of each energy transfer
rate in producing the equilibration.

\subsection{Case 1: $n_x=1$ instability at $\e/\ecz=9$}

\begin{figure*}
\centering
\includegraphics[width=.45\textwidth]{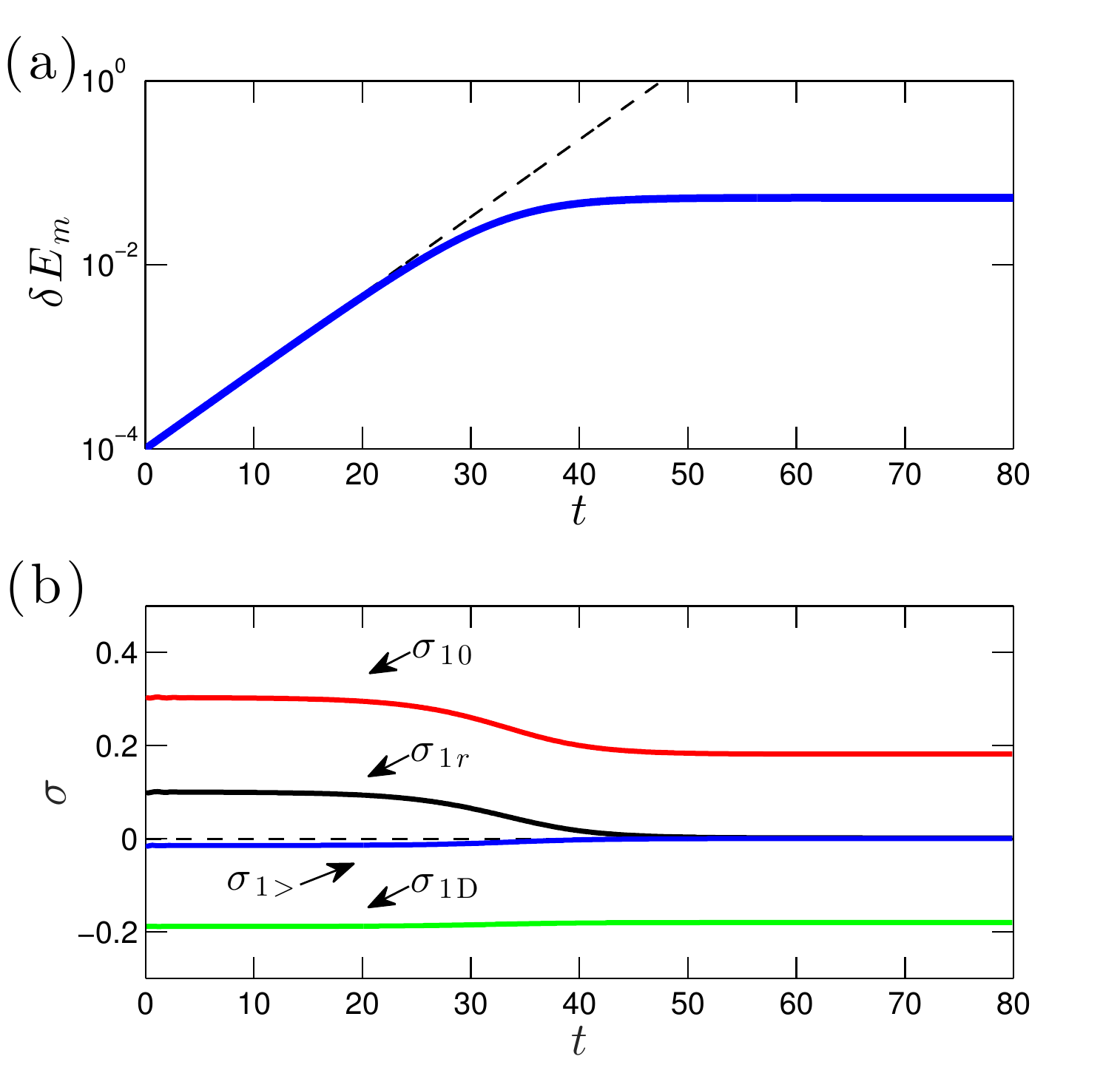}
\vspace{-1em}
\caption{(a) Evolution of the disturbance energy, $\d E_m$, of the deviation of the large-scale flow from its zonal equilibrium state at $\e/\ecz = 9$ with equilibrium vorticity $Z^e$. The S3T equilibrium  is initially perturbed with the unstable $n_x=1$ S3T eigenfunction shown in Fig.~\ref{fig:spec_comb_9p00}d. Initially the deviation grows at the predicted exponential growth rate of the eigenfunction (dashed) and the equilibration of this instability produces asymptotically the stationary state shown in Fig.~\ref{fig:flowstate_e9p00_nx1}a,b comprising a jet with a finite amplitude embedded wave. (b) Evolution of the energetics of the $k_x=1$ component of the flow. Shown are the contribution to the instantaneous growth rate of $k_x=1$ by energy transferred from the mean flow ($\s_{10}$), from the small scales ($\s_{1>}$) and that lost to dissipation ($\s_{1\textrm{D}}$). Also shown is the resulting instantaneous growth rate: $\s_{1r}=\s_{10}+\s_{1>}+\s_{1\textrm{D}}$, which necessarily vanishes as equilibration is approached. The S3T instability is supported  in this case solely from energy transferred to $k_x=1$ from $U^0$ and equilibration is achieved by reducing this transfer.} \label{fig:dEm_e9p00}
%
\vspace{1em}%
\centerline{\includegraphics[width=0.63\textwidth]{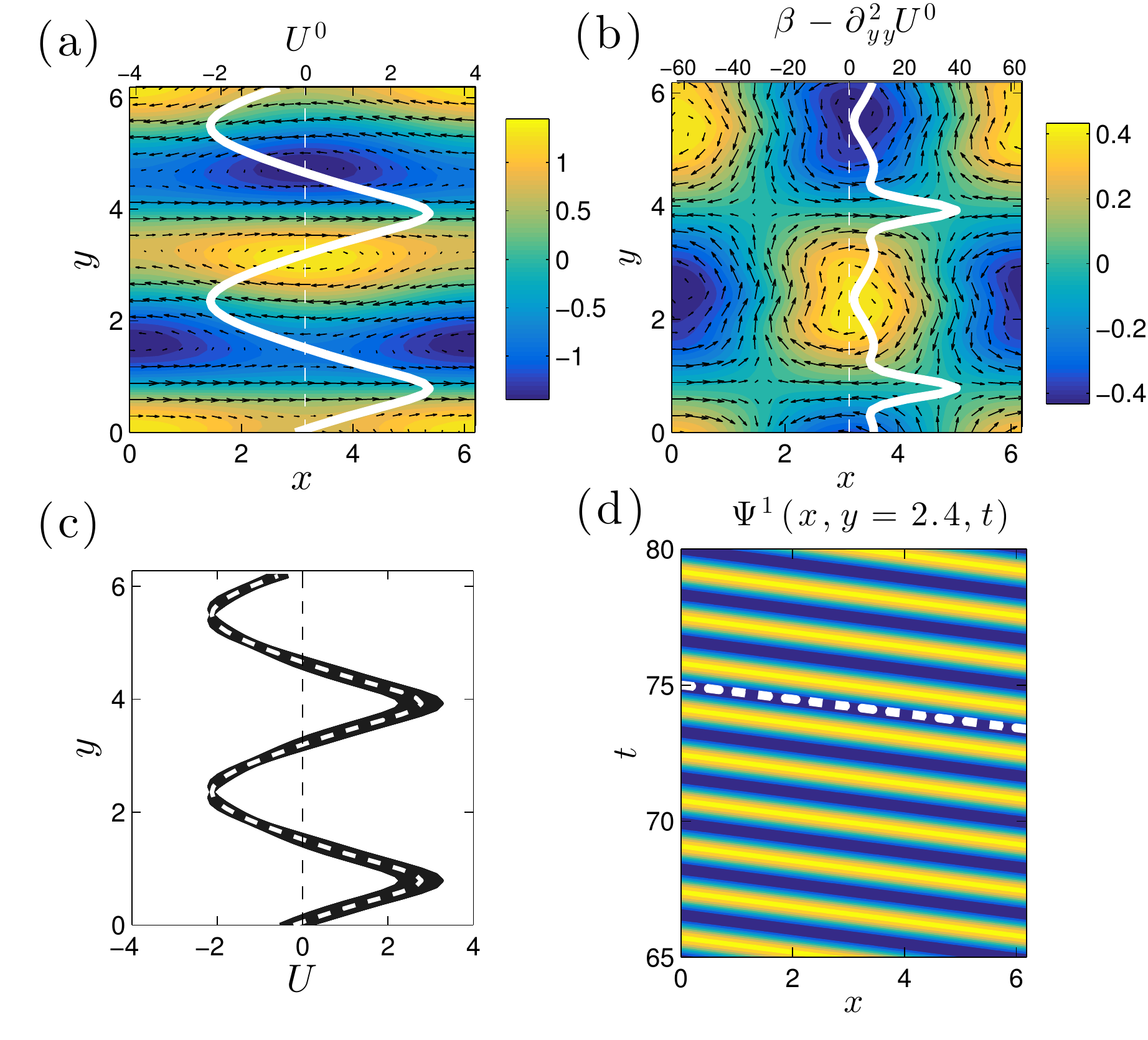}}
\caption{(a) Mean flow streamfunction, $\Psi$, at $t=80$, and the velocity field 
of the $k_x=0$ and $k_x=1$ components of the equilibrium at $\e/\ecz=9$ resulting from  equilibration of the $n_x=1$ instability.  Also shown in white is $U^0$. The equilibrium consists of  a jet 
and a traveling wave that has no critical layer in the flow as it travels retrograde with respect to the minimum jet velocity. (b) The wave component of the flow, $\Psi^1$, and its associated velocity field. The wave propagates in the retrograde part of the jet where the potential vorticity gradient, $\beta - \partial^2_{y}U^0$ (shown in white) has a small and nearly constant positive value. (c) Variation of the zonal flow velocity, $U$, with $y$ at equilibrium  at different $x$ sections. Also shown is $U^0$ (dashed line) which is nearly identical to the unstable  S3T jet $U^e$. (d) Hovm\"oller diagram of $\Psi^1$ at the location of the minimum of $U^0$, $y=2.4$. The phase velocity of the equilibrated wave is equal to the phase speed  (indicated with the dashed line)  of the most unstable S3T eigenfunction shown in Fig.~\ref{fig:spec_comb_9p00}d.}
\label{fig:flowstate_e9p00_nx1}
\end{figure*}

Consider first  the equilibration of the $n_x=1$ instability at
$\e/\ecz=9$ by first
imposing  on the jet  equilibrium the most unstable S3T
$n_x=1$ eigenfunction at small amplitude, in order to initiate its
exponential growth phase. Evolution of the energy of the $Z^1$
component of the flow  as a function of time, shown in
Fig.~\ref{fig:dEm_e9p00}a, confirms the accuracy of our methods for
determining  the structure and the growth rate of the maximally growing
S3T eigenfunction of the  jet equilibrium. The contribution of each of
the growth rates associated with~\eqref{eq:e1a}-\eqref{eq:e1d} to the
total normalized energy growth rate  of the $k_x=1$ component of the
flow, $\df E_1/\df t$, is shown in Fig.~\ref{fig:dEm_e9p00}b.  As discussed
earlier, the S3T instability is due to the transfer of energy from the
zonal flow and the equilibration is seen to be achieved by reducing the
transfer of the energy from the mean flow to the $k_x=1$ component by
reducing the tilt of the non-zonal component of the flow. The Reynolds
stress contribution remains approximately energetically neutral. The flow eventually
equilibrates to a nearly zonal configuration which is very close to the
initial  jet, as shown in Fig.~\ref{fig:flowstate_e9p00_nx1}c. The
equilibrium state while nearly zonal contains an embedded traveling wave
(cf.~Fig.~\ref{fig:flowstate_e9p00_nx1}a,b). This wave propagates
westward with phase speed indistinguishable from that of the unstable
$n_x=1$ S3T eigenfunction, as can be seen in the Hovm\"oller diagram
of $\Psi^1$, shown in
Fig.~\ref{fig:flowstate_e9p00_nx1}d.
 The PV gradient of the equilibrated jet
$\beta - \partial^2_{y} U^0$ is everywhere positive and the wave
propagates in the retrograde part of jet where the PV gradient is close
to uniform. Also the structure of the non-zonal component of the
equilibrated flow is very close to the structure of the most unstable
eigenfunction, as seen by comparing Fig.~\ref{fig:spec_comb_9p00}d  with
Fig.~\ref{fig:flowstate_e9p00_nx1}b. This equilibrated state is robustly
attracting. When the unstable jet is perturbed with random high
amplitude perturbations the unstable S3T jet is attracted to the same
equilibrium. Mixed S3T equilibria of similar form have been
found as statistical equilibria of the full nonlinear equations~
\citep{Bakas-Ioannou-2013-prl,Bakas-Ioannou-2014-jfm}.

\begin{figure*}
\centering
\includegraphics[width=0.4\textwidth]{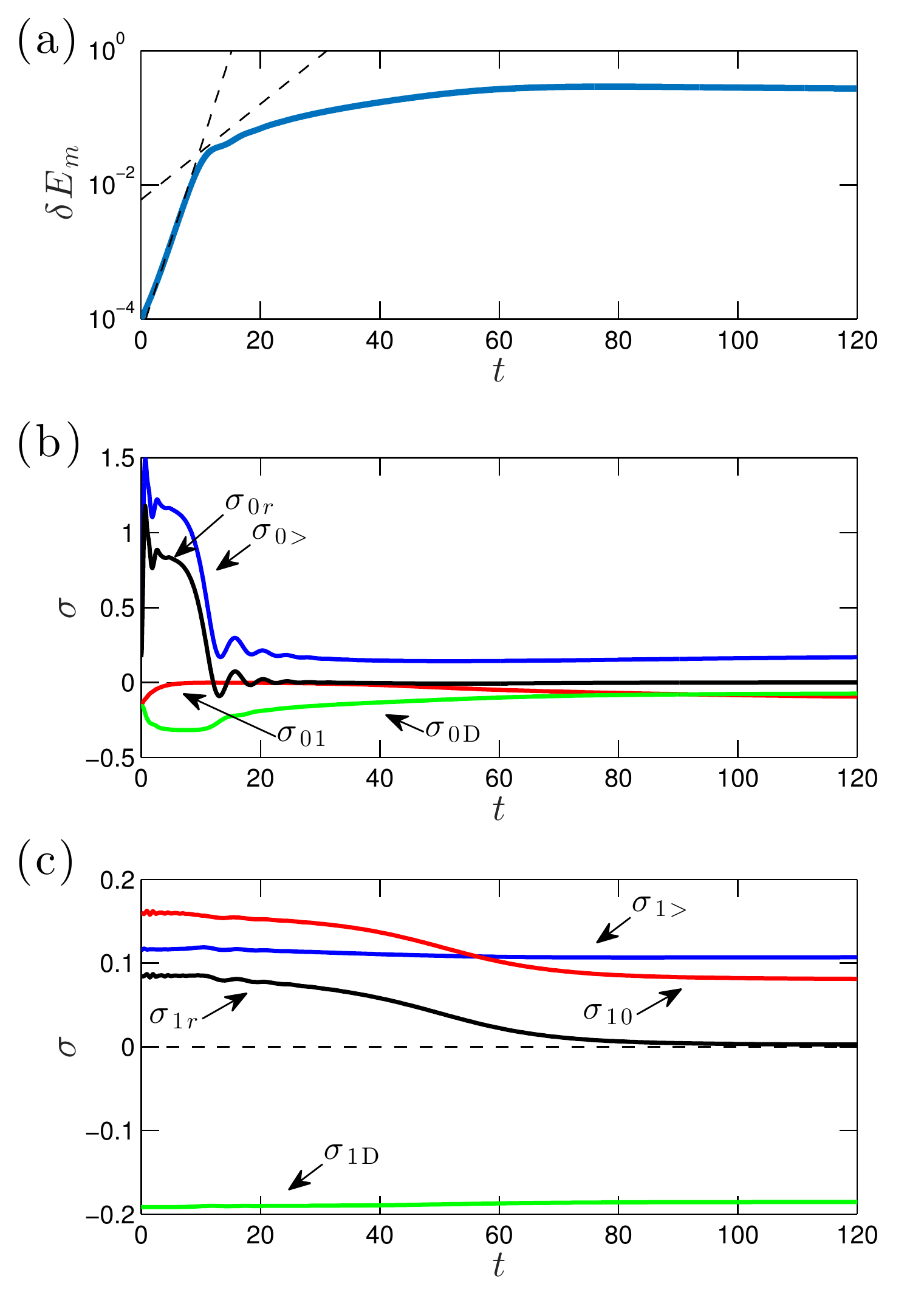}
\vspace{-1em}
\caption{Evolution of the disturbance energy, $\d E_m$, associated with the deviation of the large-scale flow $Z-Z^e$, where $Z^e$ is the zonal equilibrium vorticity, at $\e/\ecz=13.65$. 
The S3T equilibrium is initially perturbed with  the unstable $n_x=0$ and $n_x=1$ S3T eigenfunctions at small but equal amplitude. The $n_x=0$ eigenfunction grows at $\s_r=0.324$; the $n_x=1$ at $\s_r=0.083$ (both indicated  with dashed lines). Energy grows at first at the rate of the $n_x=0$ instability, up to $t\approx12$,  at which time the equilibration of  $Z^0$ is established. The equilibration of $Z^1$ is not established until $t \approx 60$. (b) Evolution of the energetics of the $Z^0$. Shown are the contribution to the instantaneous growth rate of  $Z^0-Z^e$ from energy transferred: from $Z^1$ ($\s_{01}$), from the small scales ($\s_{0>}$), and that lost to dissipation ($\s_{0\textrm{D}}$). Also shown is the actual instantaneous  growth rate, $\s_{0r}$, which vanishes at equilibration. The $n_x=0$ S3T instability is supported by the transfer of energy  from the small scales and equilibration is achieved rapidly by reducing this transfer. (c) Same as (b) but for $Z^1$. Shown are the transfer rate from $Z^0$ ($\s_{10}$), from the small scales ($\s_{1>}$) and the energy dissipation rate $\s_{1\textrm{D}}$. The $n_x=1$ instability is supported by both transfer from $Z^0$ and from small scales and the equilibration is established by reducing the transfer from $Z^0$.} \label{fig:dEm_e13p65}

\vspace{1em}

\centerline{\includegraphics[width=.55\textwidth]{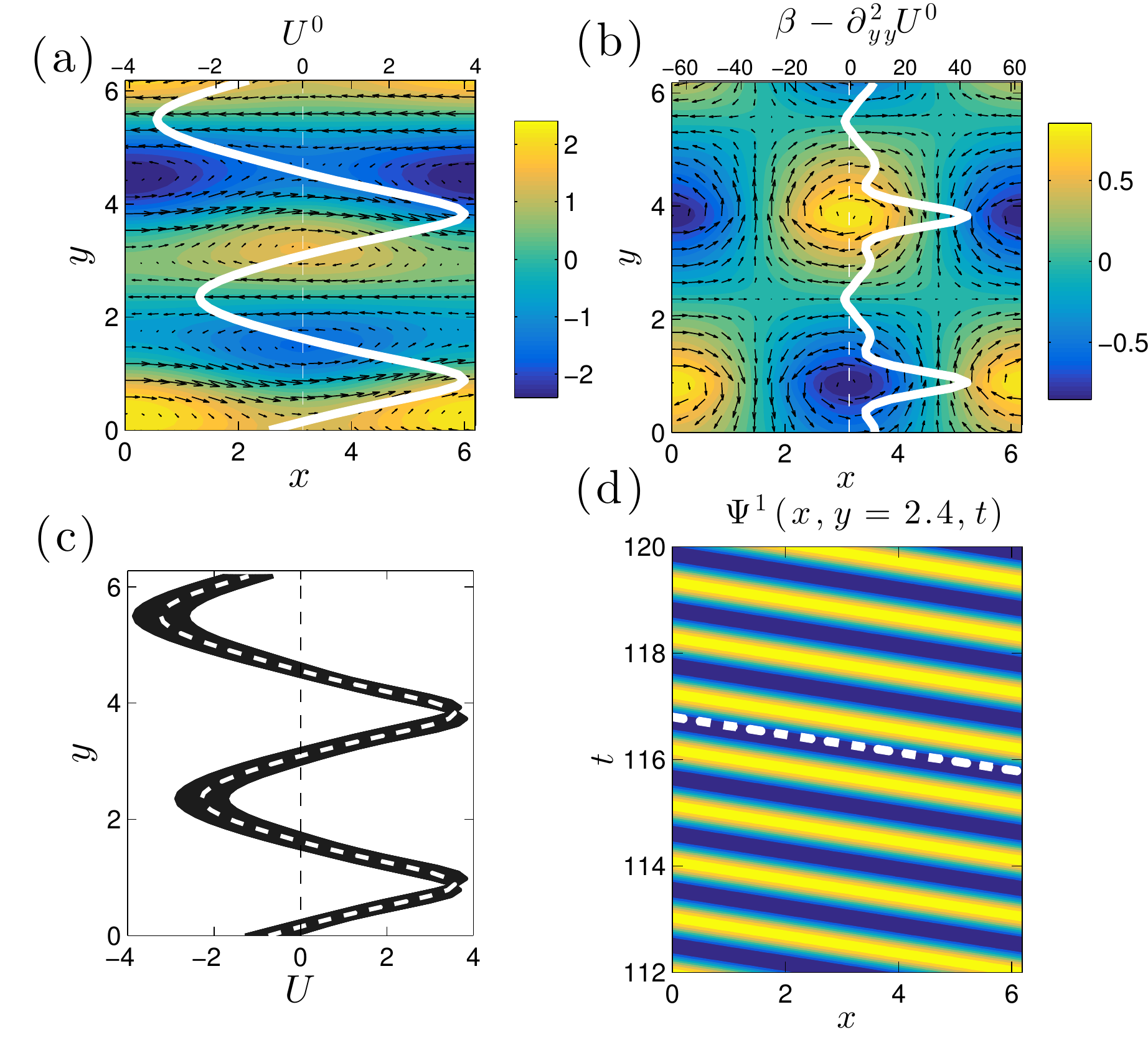}}\vspace{-1em}
\caption{(a) Mean flow streamfunction, $\Psi$, at $t=120$, and the velocity field 
of the $k_x=0$ and $k_x=1$ components of the equilibrium at $\e/\ecz=13.65$ resulting from  equilibration of both the $n_x=0$ and $n_x=1$ instabilities. Also shown in white is $U^0$. The equilibrium consists of a jet and a traveling wave that has no critical layer in the flow as it travels retrograde with respect to all $U^0$. (b) The wave component of the flow, $\Psi^1$, and its associated velocity field. The wave is trapped in the prograde part of the flow where the potential vorticity gradient (shown in white) is large. (c) Variation of the zonal flow velocity, $U$, with $y$ at equilibrium at different $x$ sections. Also shown is $U^0$ (dashed line) which is nearly identical to the unstable S3T jet $U^e$. The equilibrated jet is asymmetric. (d) Hovm\"oller diagram of $\Psi^1$ at the location of the zero of $U^0$, $y=1.6$. The phase velocity of the equilibrated wave is equal to the phase speed (indicated with the dashed line)  of the most unstable S3T eigenfunction shown in Fig.~\ref{fig:spec_comb_13p65}d.} \label{fig:flowstate_e13p65_nx1}
\end{figure*}

\subsection{Case 2: $n_x=1$ instability at $\e/\ecz=13.65$}

The equilibration of the jet at $\e/\ecz=13.65$ involves the simultaneous
equilibration of two S3T instabilities, of the powerful $n_x=0$ jet
instability that grows initially at the rate $\s_r=0.324$ and of the weaker 
$n_x=1$ instability that grows initially at rate $\s_r=0.083$. We impose on the
equilibrium  the most unstable S3T $n_x=0$ and $n_x=1$ eigenfunctions at
small but equal amplitudes, in order to initiate their exponential
growth phases. The evolution of the energy of the $Z-Z^e$ component
of the flow as a function of time (cf.~Fig.~\ref{fig:dEm_e13p65}a)
shows initial growth at the rate of the faster $n_x=0$ instability. The 
equilibration process for the $n_x=0$ instability is shown in
Fig.~\ref{fig:dEm_e13p65}b, and the equilibration of the $n_x=1$
instability in Fig.~\ref{fig:dEm_e13p65}c. The $n_x=0$ instability is
supported by the transfer of energy to the $k_x=0$ component from the
small scales ($\sigma_{0>} > 0$) as is the equilibrated jet.  The
equilibration of this instability proceeds rapidly and is  enforced by
reduction of the $\sigma_{0>}$, i.e. the transfer of energy from the
small scales. During the equilibration process there is a pronounced
transient enhancement of the transfer rate to the mean flow by the
eddies. This leads to the equilibrated jet shown in
Fig.~\ref{fig:flowstate_e13p65_nx1}a,c which has $5\%$ greater energy
than the original S3T unstable equilibrium jet. The equilibrated jet is
asymmetric with enhanced power at $n_y=1$. (In this case the
unstable $n_y=2$ jet did not merge with the $n_y=1$ jet to 
form a jet with a single jet structure.)
During the equilibration process $\sigma_{01}$ is always negative,
indicating  continual transfer of mean jet energy supporting the
$n_x=1$ perturbation. 
The equilibration of the $n_x=1$ wave is slower and
proceeds  in this example, in which the jets did not merge,
independently of  evolution  of the $n_x=0$ instability.  The wave is
supported by transfer of energy from the small scales and from
transfer of energy from the mean flow. The former remained unaffected
during the equilibration process and equilibration is achieved by
reduction of the transfer  from the mean flow, $\s_{10}$. The PV gradient of the mean flow,
$\beta - \partial^2_{y} U^0$, shown in~Fig.~\ref{fig:flowstate_e13p65_nx1}b 
is positive almost everywhere and the wave is trapped at the prograde part of 
the jet. As in the case with $\e/\ecz=9$, the wave propagates at the speed of 
the S3T eigenfunction (cf.~Fig.~\ref{fig:flowstate_e13p65_nx1}d).

\section{Discussion}

\subsection{Correspondence between the S3T dynamics~\eqref{eq:s3t} and the projected S3T dynamics~\eqref{eq:s3tf}}

\begin{figure*}
\centerline{\includegraphics[width=0.55\textwidth]{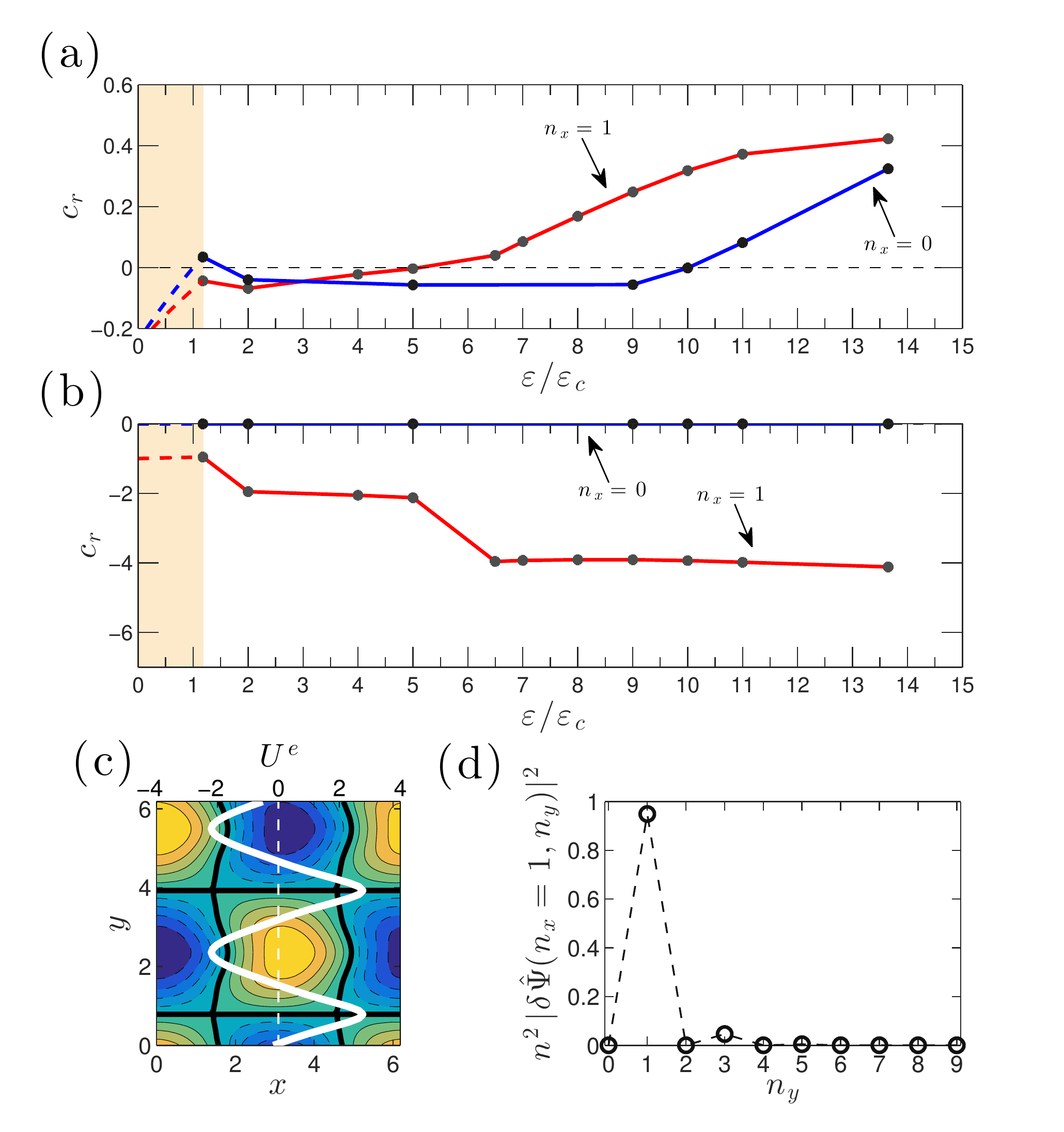}}
\vspace{-5mm}
\caption{(a)-(b): The stability of the jet equilibria in the S3T formulation~\eqref{eq:s3t}. The corresponding stability properties of the projected S3T system
are shown in  Fig.~\ref{fig:S3Tgrowth}a-b. (c):~Contour plot  of the streamfunction of the most unstable non-zonal $n_x=1$ S3T mean flow eigenfunction of the $n_y=2$ jet equilibrium at $\e/\ecz=9$ with growth rate $\s_r=0.248$ and phase speed $c_r=-3.91$. The equilibrium jet plotted in solid white. Positive (negative) contours are shown with solid (dashed) lines and the zero contour with thick solid line. (b):~The energy power spectrum of the mean flow eigenfunction. The jet is stable to $n_x=0$ perturbations but unstable to $n_x=1$ perturbations and the most unstable $n_x=1$ eigenfunction is Bloch $q_y=1$ with power at $n_y=1$.} \label{fig:S3TgrowthNoNPK_eig9}
\vspace{5em}
\centerline{\includegraphics[width=0.90\textwidth]{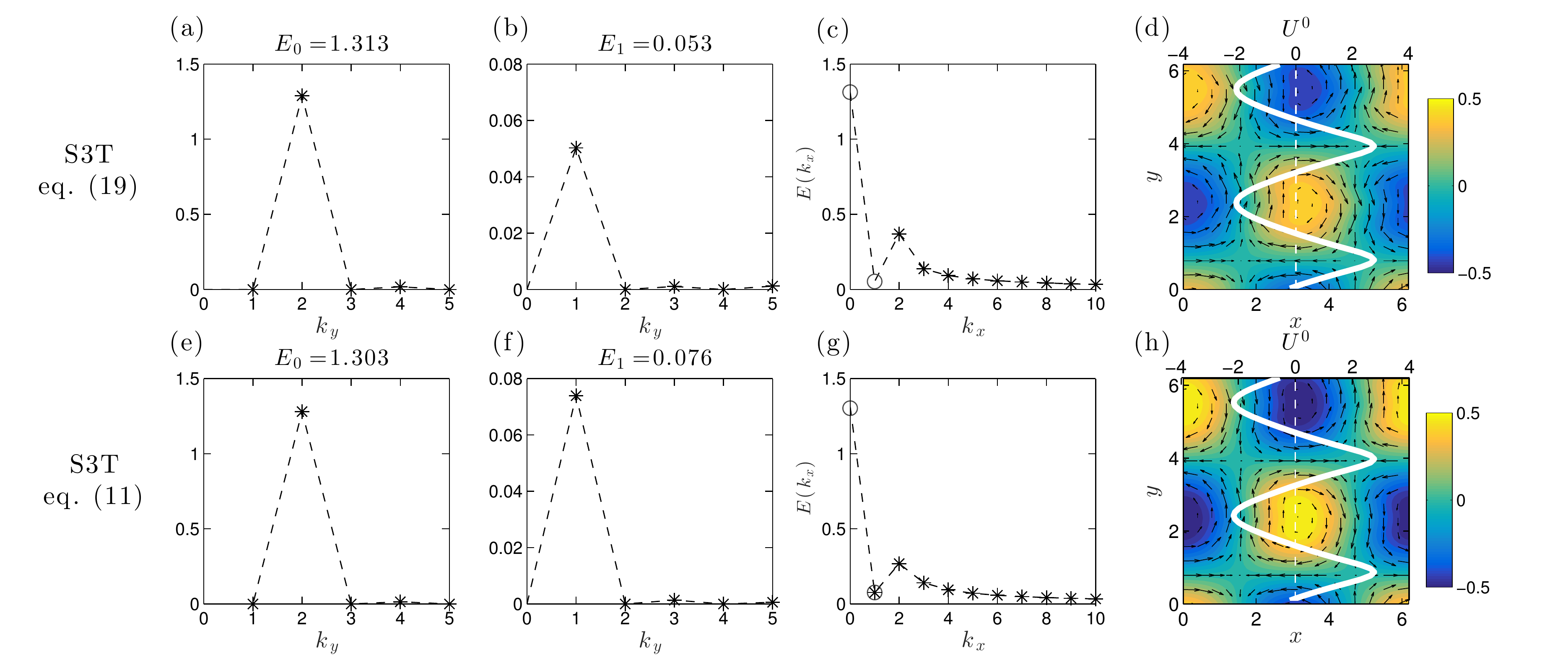}}
\vspace{-5mm}
\caption{Comparison of the flows resulting from  equilibration of the $n_x=1$ S3T instabilities of
the projected S3T system~\eqref{eq:s3t_Pk} (panels (a)-(d)) and the 
S3T system~\eqref{eq:s3t} (panels (e)-(h)) at $\e/\e_c=9$. Shown are: The $k_y$ energy spectrum of the mean flow ($k_x=0$) (first column) and of the $k_x=1$ (second column), the $k_x$ energy spectrum  of both the coherent flow components ($|k_x|\le 1$, shown with circles) and of the incoherent flow components $|k_x|>1$ (shown with asterisks) (third row). Snapshots of the mean jet (thick line) and contour plot of the streamfunction of the $k_x=1$ wave component are shown in the figures of the fourth column.} \label{fig:S3T_PKvsNikos}
\end{figure*}

Stability of a two jet state to jet/wave perturbations  in the projected S3T formulation~\eqref{eq:s3tf}
is shown in Fig.~\ref{fig:S3Tgrowth}.
For parameters  for which the base state becomes 
unstable to non-zonal large-scale perturbations
this base state  transitions to a new equilibrium in which the jet
coexists with a coherent wave.  The stability calculation, its  energetics  and equilibration process 
are studied  in the framework of  the projected S3T equations~\eqref{eq:s3tf}, which allows 
a clear separation between the contribution of the coherent jet interaction and that of   the incoherent eddies to the instability and equilibration processes. 
This stability analysis using the projected S3T system produces essentially the same results as were obtained using 
the S3T system~\eqref{eq:s3t} 
 (compare Fig.~\ref{fig:S3Tgrowth} with Fig.~\ref{fig:S3TgrowthNoNPK_eig9}a-b and  Fig.~\ref{fig:eig_nx1_spec}e,f  with
  Fig.~\ref{fig:S3TgrowthNoNPK_eig9}c-d). The equilibrated states produced by  these two S3T systems are also very similar (cf.~Fig.~\ref{fig:S3T_PKvsNikos}).

\begin{figure*}
\vspace{3em}\centerline{\includegraphics[width=0.90\textwidth]{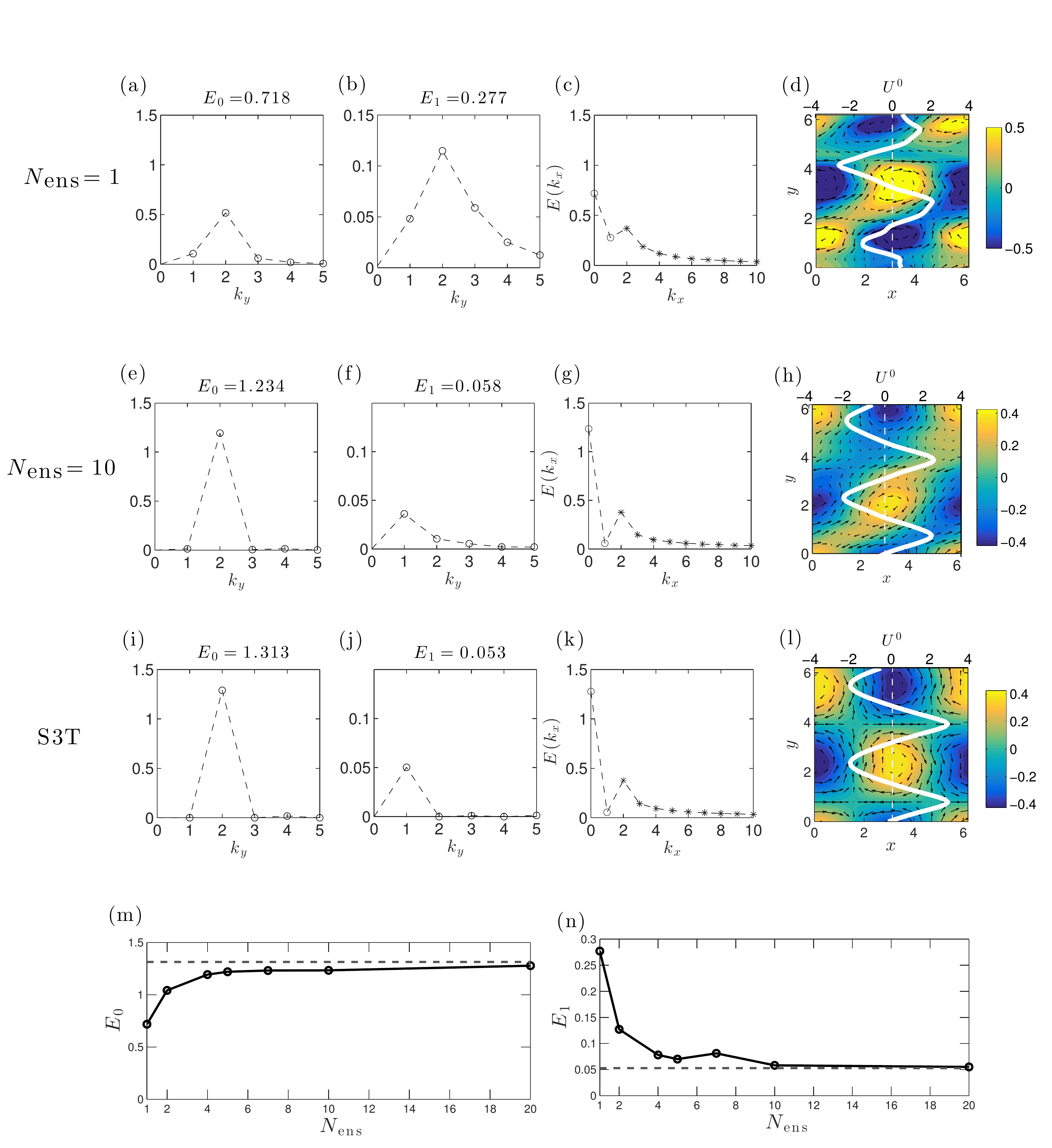}}\vspace{2em}
\caption{(a)-(l): Structure of the mean flow in ensemble QL simulations as the number of ensemble members, $N_{\textrm{ens}}$, increases. Shown are $N_{\textrm{ens}}=1$ (first row, (a)-(d)) and  $N_{\textrm{ens}}=10$  (second row, (e)-(h))  at $\e/\e_c=9$. Also shown is the equilibrated mean flow from the S3T simulation (third row, (i)-(l)) at the same forcing amplitude (cf.~Fig.~\ref{fig:flowstate_e9p00_nx1}).  Shown are: The $k_y$ energy spectrum of the mean flow ($k_x=0$) (first column) and of the $k_x=1$ (second column), the $k_x$ energy spectrum  of both the coherent flow components ($|k_x|\le 1$) and of the incoherent flow components  $|k_x|>1$ (third row). Snapshots of the mean jet (thick line) and contour plot of the streamfunction of the $k_x=1$ wave component is shown in the figures of the fourth column. (m) The approach of the energy of the zonal flow, $E_0$ , obtained in ensemble QL simulations as the number of ensemble members, $N_{\textrm{ens}}$ increases  for the case $\e/\e_c=9$. Dashed line marks the S3T prediction. (n): The approach of the wave component energy $E_1$ to the S3T predictions (marked by dashed line). Convergence is achieved with $N_{\textrm{ens}}=10$. Note that for small $N_{\textrm{ens}}$ fluctuations result in enhanced excitation of the $k_x=1$ component. Parameters as in previous simulations.}
\label{fig:ensembleQL}
\end{figure*}

\begin{figure*}

\centerline{\includegraphics[width=0.95\textwidth]{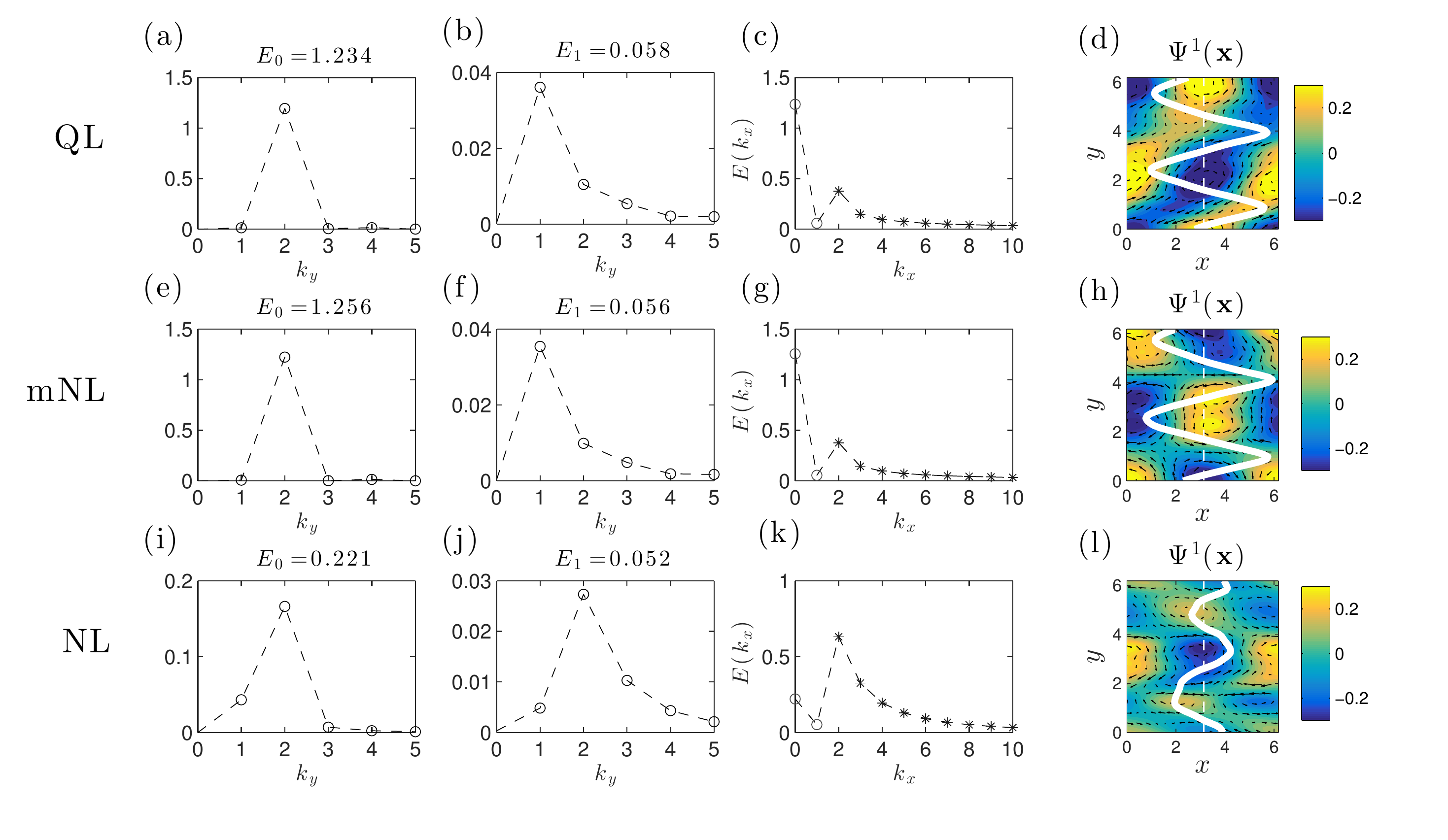}}
\caption{QL simulations (panels (a)-(d)), modified NL simulations (panels (e)-(h)) in which only the term $(I-P_K )\[\nablav \bcdot\(  \uv' \zeta' \)\]$ is removed, and unmodified ensemble NL simulations (panels (i)-(l)), all with 10 ensemble members  at $\e/\e_c=9$. Shown are: the $k_y$ energy spectrum of the mean flow ($k_x=0$) (first column), of the $k_x=1$ component (second column), the $k_x$ energy spectrum, of both the coherent flow components ($|k_x|\le 1$) and of the incoherent flow components $k_x=1$ (third column). Snapshots of the mean jet (thick line) and contour plot of the streamfunction of the wave component $k_x=1$ is shown in
the figures of the fourth column. Parameters as in the other simulations. Note that the eddy--eddy interactions in  NL  lead  in this particular example
to  an appreciable departure from the QL spectrum of the eddy components, 
i.e. wavenumbers  $k_x\ge 2$ (cf. panels (k) and (c)).   This is caused because all the eddy components   $k_x\ge 2$ 
are externally strongly forced (the dissipation has an e-folding of $40~\rm d$). This figure shows that the eddy--eddy nonlinearity
is the source of the difference between the ensemble QL and the ensemble NL simulations.} \label{fig:eqlnl}
\end{figure*}

\subsection{Reflection of ideal S3T dynamics in QL simulations}

The ideal S3T equilibrium jet and jet/wave states that we have obtained
are imperfectly reflected in single realizations of the flow
because  fluctuations may obscure the underlying S3T equilibrium 
(cf.~\cite{Farrell-Ioannou-2003-structural,Farrell-Ioannou-2015-book}). 
The infinite ensemble ideal incorporated in the S3T dynamics can be approached in the QL (governed by~\eqref{eq:PKQL}) 
by introducing in the equation for the coherent flow an ensemble mean Reynolds stress  obtained from a number of independent integrations of the QL 
eddy equations with different forcing realizations.

Consider for example the jet/wave S3T regime  at $\e/\e_c=9$ shown in  Fig.~\ref{fig:flowstate_e9p00_nx1}.
The energy of the $k_x=0$ component of the coherent flow is $E_0=1.3 $ and of the $k_x=1$ component,
which is predominantly a $k_y=1$ wave, is $E_1=0.05$. 
In Fig.~\ref{fig:ensembleQL}a-l is shown the approach of the QL dynamics to this ideal S3T equilibrium as a function of the number of ensemble members, $N_{\textrm{ens}}$, using as diagnostics the structure, indicated by snapshots, of the coherent flow and the energy spectrum. Convergence of the energy of the QL coherent flow components to that of the S3T as $N_{\textrm{ens}}$ increases is shown in
Fig.~\ref{fig:ensembleQL}m-n. These ensemble QL simulations were performed
by  introducing the mean Reynolds stress divergence  obtained from $N_{\textrm{ens}}$ independent
simulations of~\eqref{eq:PQL}, all with the same large-scale flow,  obtained from a single mean QL equation~\eqref{eq:MQL}.
Convergence to the S3T state is closely approached with $N_{\textrm{ens}}=10$.  In simulations with 
a smaller number of ensemble members the ensemble QL supports an irregular weaker $k_y=2$ jet 
and a stronger $k_x=1$ coherent flow, which is concentrated at $k_y=2$ rather than
at $k_y=1$ as predicted in the S3T  (cf. Fig.~\ref{fig:ensembleQL}b). 
As the number of ensemble members increases the jet is more coherently 
forced and the ideal S3T $k_x=1$ component, which was previously masked by fluctuations at $k_y=2$, is revealed.
Also note that in these QL simulations there are no eddy--eddy interactions and also no direct stochastic
forcing of the coherent flow components and consequently their emergence does not result from cascades but from the structural instability mechanisms revealed by S3T.     

 Both S3T and ensemble simulations isolate and clearly reveal the mechanism by which a portion of the incoherent turbulence is systematically organized by large-scale waves to enhance the organizing wave. However, as in simulation studies revealing this mechanism at work in baroclinic turbulence \citep{Cai-Mak-1990, Robinson-1991b}, the large-scale wave retains a substantial incoherent 
component in individual realizations.  This is expected in the strongly turbulent atmosphere considering that even stationary waves at planetary scale which are strongly forced by topography are revealed clearly only in seasonal average ensembles.

\subsection{Reflection of ideal S3T dynamics in NL simulations}

Consider now the reflection of the S3T  jet/wave regime in NL and ensemble NL simulations. 
Ensemble simulations of the NL system~\eqref{eq:PKNL}  were performed by
introducing in the mean equations~\eqref{eq:PKMNL} 
the ensemble average of $P_K \( \uv' \bcdot \nablav \zeta' \)$ and
$P_K\( \Uv \bcdot \nablav \z' + \uv' \bcdot \nablav Z \)$ obtained from $N_{\textrm{ens}}$ independent simulations of the perturbation NL equations~\eqref{eq:PKPNL}
all with the same large-scale flow.
The corresponding results of the ensemble QL simulation (cf.~Fig.~\ref{fig:eqlnl}a-d)
differ from those of the ensemble NL simulation. The
nonlinear term  $(I-P_K )\(  \uv' \bcdot \nablav \zeta' \)$ is responsible for the difference between the 
NL and QL ensemble simulations, as shown in Fig.~\ref{fig:eqlnl}e-h. In this figure an ensemble integration of the NL equations with
this term  absent is shown to produce results that are very close to the QL results.

\begin{figure*}
\vspace{2em}\centerline{\includegraphics[width=0.95\textwidth]{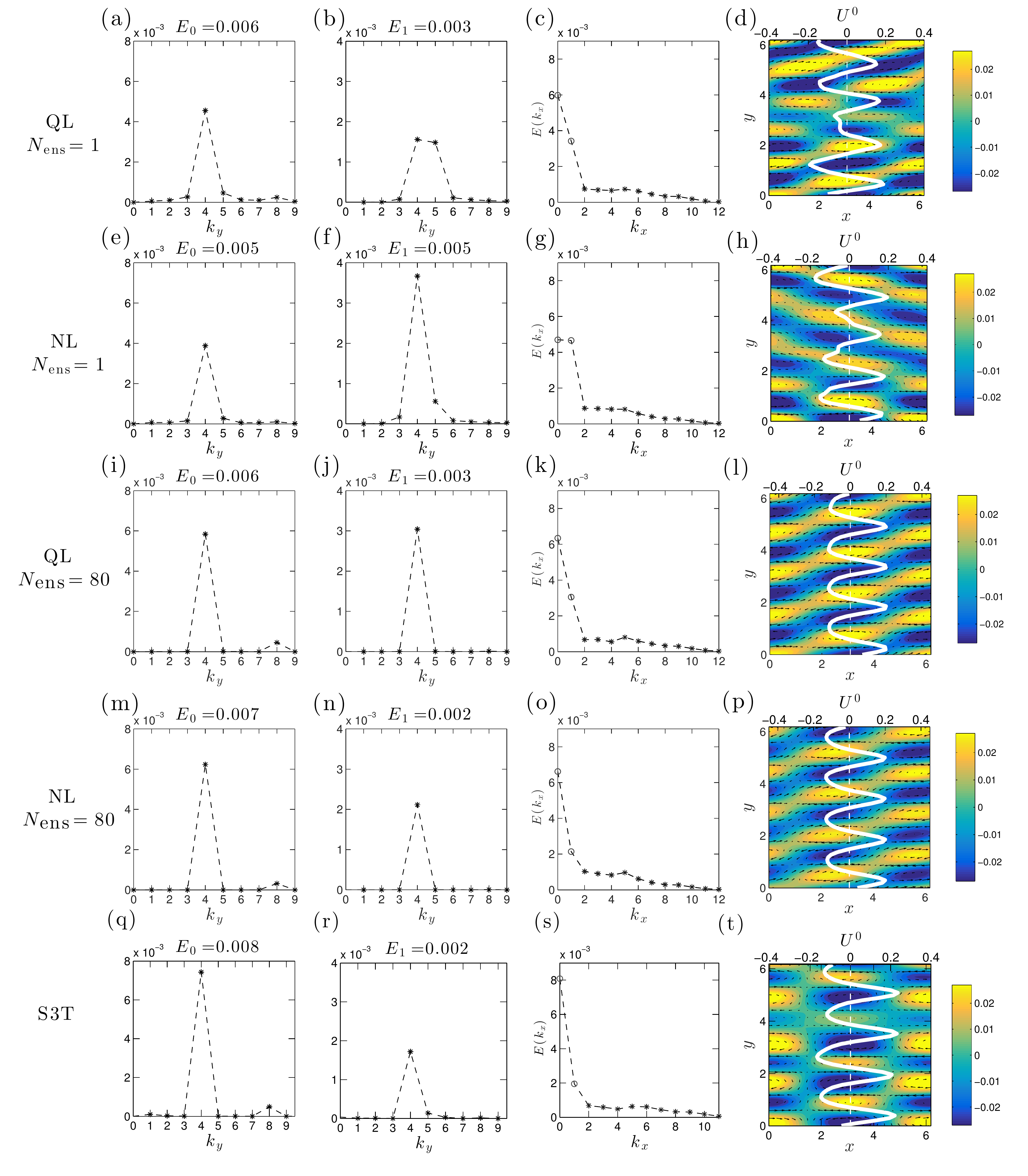}}
\vspace{2em}\caption{Comparison among QL and NL simulations with ensemble members $N_{\rm ens}=1$ 
 (panels (a)-(h)), $N_{\rm ens}=80$  (panels (i)-(p)) and S3T simulation (panels (q)-(t)). The S3T state in this example predicts
a jet component and a $k_x=1$ component and this is  reflected  in both  QL and NL ensemble simulations as revealed by the
 $k_y$ energy spectrum of the respective mean flow ($k_x=0$) (first column),
 the $k_x=1$ component (second column), and 
 the $k_x$ energy spectrum
 (third column). Snapshots of the mean flow (thick line) and contour plot of the streamfunction of the wave component $k_x=1$ are shown in panels  (d), (h), (l), (p) and (t) (fourth column). Simulations
 with isotropic forcing at total wavenumber $K_f=10$ (cf.~\eqref{eq:finite_ring}) with $\e=4.2\times10^{-4}$, linear damping coefficient $r=0.01$ and $\nu=0$.}
  \label{fig:nlql_agree}
\end{figure*}

\begin{figure*}
\vspace{2em}\centerline{\includegraphics[width=0.95\textwidth]{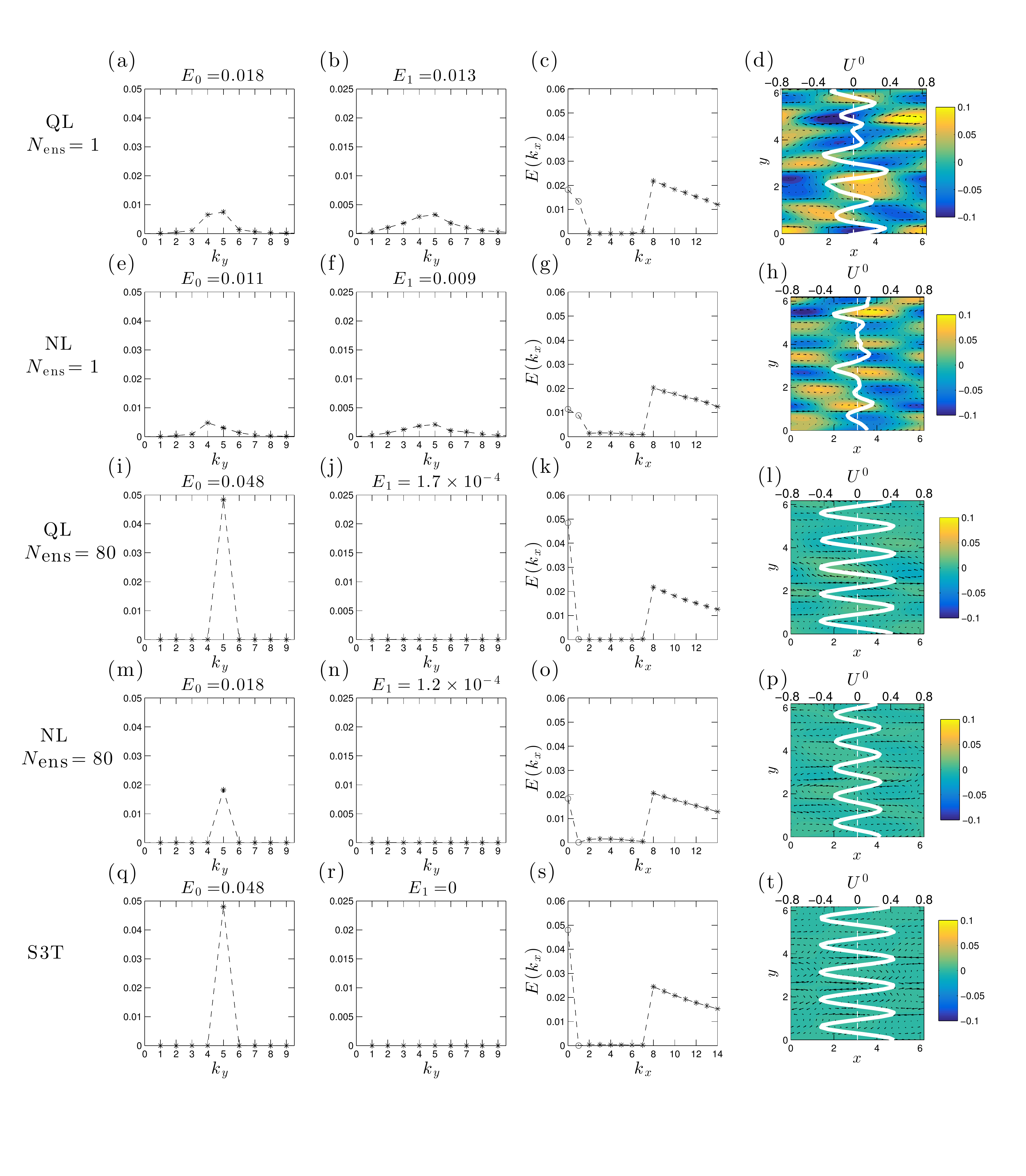}}
\caption{Comparison among QL and NL simulations with ensemble members $N_{\rm ens}=1$ 
 (panels (a)-(h)), $N_{\rm ens}=80$  (panels (i)-(p)) and S3T simulation (panels (q)-(t)). The S3T state in this example predicts
only a jet component and no $k_x=1$ component and this  is  reflected  in both  QL and NL ensemble simulations  as revealed by the
 $k_y$ energy spectrum of the respective mean flow ($k_x=0$) (first column),
 the $k_x=1$ component (second column), and 
 the $k_x$ energy spectrum (third column). Snapshots of the mean flow (thick line) and contour plot of the streamfunction of the wave component $k_x=1$ are shown in panels 
 (d), (h), (l), (p) and (t) (fourth column). Simulations
 with the anisotropic forcing spectrum~\eqref{eq:spec_NIF}  at $\e=0.830$  with zonal wavenumbers $8\le|k_x|\le14$ being forced.
 Mean linear damping coefficient is $\rM=0.1$ and linear damping coefficient of the incoherent flow is $r=1.5$ and $\nu=0$.}  \label{fig:nlql_agree1}
\end{figure*}

When all waves with $|k_x|\ge 2$ are forced equally, as 
in the S3T examples discussed above, the eddy--eddy interactions are strong in the corresponding NL 
resulting in a substantial modification of the spectrum of the eddy motions which is not reflected
in the S3T.
To obtain correspondence an effective stochastic forcing which parameterizes the absent eddy--eddy interactions
is required in S3T \citep{Constantinou-etal-2014}.

Alternatively,  when the term    $(I-P_K )\[\nablav \bcdot\(  \uv'   \zeta' \)\]$
is suppressed by choosing low forcing excitation, which results in weak modification
of  the spectrum of the incoherent component, agreement between 
NL and QL simulations is obtained. This is demonstrated in Fig.~\ref{fig:nlql_agree}, in which 
we show results obtained with an approximate small-scale isotropic ring forcing:
\be
\hat{Q}(\kv)=\left\{
	\begin{array}{ll}
		\dfrac{4\pi }{\log{\(\frac{K_f+\d K_f}{K_f-\d K_f}\)}} & \mbox{if }~ K_f- \d K_f \leq k \leq K_f+ \d K_f \ ,\\
		0 & \mbox{if } ~\left|\bit k-K_f\right|>\d K_f  \text{ or } |k_x|\le 1\ ,
	\end{array}
\right. \label{eq:finite_ring}
\ee
with  $K_f=10$, $\d K_f=1$ and $r=0.01$, $\nu=0$ , as in \cite{Bakas-Ioannou-2014-jfm}.
With these parameters 
the S3T zonal jet equilibrium  is stable 
to jet perturbations and unstable to $n_x=1$ wave perturbations
and the resulting equilibrium state in NL has a wave $k_x=1$ component in agreement
with S3T predictions. 
Also the  energetics of the mechanism of  destabilization of the  external $n_x=1$ Rossby wave is 
partitioned between coherent and incoherent sources consistent 
with the mechanism described in the previous sections. 

It could be maintained that because  isotropic ring forcing  suppresses  eddy--eddy interactions,
the  agreement between S3T and NL  should be expected (cf.~\citet{Bakas-etal-2015},~Appendix ~C).  
This property follows from the fact that a barotropic fluid excited in an infinite channel with an isotropic ring forcing with spectrum $Q(\kv) \propto \delta(|\kv|-K_f)$  results in  a nonlinear solution which by itself could never give rise to a jet. The emergence of jets
under this forcing  can only result from imposition of a separate perturbation such as the  jet perturbation that results in the S3T jet instability. 
As an example closer to physical reality  consider  forcing of the form~\eqref{eq:spec_NIF} which excites the  zonal wavenumbers
$8 \le |k_x|\le 14$, modeling baroclinic energy injection, and with linear damping for the incoherent scales  
$r=1.5$ and $\rM=0.1$ for the coherent scales with corresponding e-folding times of $4\,{\rm d}$  and $60\,{\rm d}$.
For these parameters S3T theory predicts that the 5 jet equilibrium  at $\e=0.830$ is 
 stable to both jet and $n_x=1$ wave  perturbations and consequently S3T theory predicts that QL and NL simulations should show suppressed energy in the $k_x=1$ coherent wave component. 
 Good agreement between QL and NL is found in the channel as shown in Fig.~\ref{fig:nlql_agree1}. It is interesting to note 
 that while the jet in the NL simulation  has the structure predicted  by the S3T its amplitude is reduced consistent with 
 a component of the  eddy--eddy interactions   acting as diffusion on the mean jet.

\section{Conclusions\label{sec:concl}}

Large-scale coherent structures such as jets, meandering jets and waves embedded
in jets are characteristic features of turbulence in planetary
atmospheres. While conservation of energy and enstrophy in inviscid 2D
turbulence predicts spectral evolution leading to concentration of
energy at large scales, these considerations cannot predict the phase
of the spectral components and therefore can not address the central
question of the organization of the energy into specific structures
such as jets and the coherent component of
planetary scale waves. In order to study structure formation
additional aspects
of the dynamics beyond conservation principles must be incorporated in
the analysis. For this purpose SSD models have been developed and used
to study the formation of coherent structure in planetary scale
turbulence. In this work an SSD model was formulated for the purpose of
studying the regime of coexisting jets and waves. In this model a
separation   in zonal Fourier modes
 is made by projection in order to separate a coherent structure equation, in
which only the gravest zonal harmonics are retained, from the remaining
spectrum, which is assumed to be incoherent and gives 
rise to the ensemble mean second order statistics
associated with the incoherent turbulence. This second order SSD model
is closed by a stochastic forcing parameterization that accounts for
both the neglected nonlinear dynamics of the small scales as well as the
forcing maintaining the turbulence. The equation for the large scales
retains its nonlinearity and its interaction through Reynolds stress
with the perturbations.

In this model jets form as instabilities and equilibrate nonlinearly at finite
amplitude. A  stable mode  of the Rossby wave spectrum associated
with these jets is destabilized for sufficiently strong forcing by
interaction with perturbation Reynolds stresses. This destabilization is
found to have in some cases the remarkable property of resulting from
destabilization of the retrograde Rossby wave to mean jet interaction by
structural modification of this damped mode arising from its interaction
with the incoherent turbulence thereby transforming it into an unstable
mode of the mean jet. In other cases comparable contributions are found
from direct forcing by the Reynolds stresses, as in S3T instability with
projections at $K=0$, and induced jet/wave interaction, as in
traditional hydrodynamic instability. This synergistic interaction
provides a powerful new mechanism for maintaining planetary waves that
will be the subject of further investigation.

\acknowledgments
We thank Nikolaos Bakas for useful discussions on the  energetics of the equations in spectral space.  We also thank 
the anonymous reviewers for their comments that led to improvement
of the paper. N.C.C. would like to thank Prof.~Georgios Georgiou for his hospitality and support during the summer of 2015 at the University of Cyprus. B.F.F. was supported by NSF AGS-1246929. N.C.C was partially supported by the NOAA Climate and Global Change Postdoctoral Fellowship Program, administered by UCAR's Visiting Scientist Programs.


\end{document}